\documentclass[usenatbib,useAMS]{mn2e}
\usepackage{epsf}
\usepackage{epsfig}
\usepackage{amsmath,amssymb}
\usepackage{times}
\usepackage{natbib}
\usepackage{dcolumn}
\usepackage[english]{babel}
\RequirePackage{amsmath}
\RequirePackage{amssymb}        
\RequirePackage{wasysym}

\title{Abundance analyses of helium-rich subluminous B stars}

\author[Naslim, N. et al.]
{
Naslim\,N.$^1$\thanks{E-mail: nas@arm.ac.uk}, 
C.S.\,Jeffery$^1$\thanks{E-mail: csj@arm.ac.uk}, 
A.\,Ahmad$^1$, 
N.T.\,Behara$^{1,2}$ \&
T.\,\c{S}ah\`in$^{1,3}$ \\
$^1$Armagh Observatory, College Hill, Armagh BT61\,9DG, N.\,Ireland \\
$^2$Institut d'Astronomie et d'Astrophysique,
  Universit\'e Libre de Bruxelles, Belgium \\
$^3$Department of Astronomy and The W.J. McDonald Observatory, The
University of Texas at Austin,  Austin, Texas 78712, USA
}

\date{Accepted .....
      Received ..... ;
      in original form .....}

\pagerange{\pageref{firstpage}--\pageref{lastpage}}
\pubyear{2010}

\begin{document}

\maketitle

\label{firstpage}

\begin{abstract}
The connection between helium-rich hot subdwarfs of spectral types O
and B (He-sdB) has been relatively unexplored since the latter were found
in significant numbers in the 1980's. In order to explore this
connection further, we have analysed the surface composition of six 
He-sdB stars, including LB\,1766, LB\,3229, SB\,21 (= Ton-S\,137 = BPS\,29503-0009),
BPS\,22940--0009, BPS\,29496--0010, and BPS\,22956--0094. Opacity-sampled line-blanketed
model atmospheres have been used to derive atmospheric properties and 
elemental abundances. All the stars are moderately metal-poor compared
with the Sun ([Fe/H]$\approx-0.5$). Four stars are
nitrogen-rich, two of these are carbon-rich, and at least four appear
to be neon-rich.  The data are insufficient
to rule out binarity in any of the sample. The surface composition and 
locus of the N-rich He-sdBs are currently best explained by the 
merger of two helium white dwarfs, or possibly by the merger of a
helium white dwarf with a post-sdB white dwarf. C-rich He-sdBs require 
further investigation. 
\end{abstract}


\section{Introduction}
Subdwarf B stars are low-mass core-helium burning stars with extremely
thin hydrogen envelopes. They behave as helium main-sequence
stars of roughly half a solar mass. Their atmospheres are 
generally helium deficient; radiative levitation and
gravitational settling combine to make helium sink below the
hydrogen-rich surface \citep{heber86}.

However, almost 5$\%$ of the total subdwarf population comprise stars
with helium-rich atmospheres \citep{green86,ahmad06}. The optical spectra of
these stars are characterised by strong neutral helium lines and 
weak He{\sc ii} lines; they exhibit a wide range of helium abundance and
effective temperatures ($T_{\rm eff}$) similar to both sdB and sdO stars. They have
been variously classified as sdOB, sdOC and sdOD \citep{green86} stars,
but more recently as He-sdB and He-sdO stars
\citep{moehler90,ahmad04}. The spectroscopic division concerns the relative strengths of  He{\sc i}\,4471, 
He{\sc ii}\,4541 and H$\gamma$ (itself a blend of H and He{\sc
  ii}). Roughly speaking, the division occurs for stars with $T_{\rm
  eff} \approx 38\,000\,{\rm K}$ \citet{drilling03}. 

In general, He-sdB stars have lower surface gravities ($g$) than normal 
hydrogen-rich sdB stars \citep{heb09} and have spectral
characteristics intermediate between extreme helium (EHe) stars 
\citep{jeffery96a} and He-sdO stars \citep{napiwotzki08}.

Most He-sdB stars show strong nitrogen (N{\sc ii} and N{\sc iii}) lines in their
optical spectrum; these are referred to as N-rich by
\citet{drilling03}. A few {\it also} showing strong carbon (C{\sc ii} and C{\sc iii}) lines are
labelled C-rich.

The question posed by these stars is that of their evolutionary
status.  
Do He-sdB and He-sdO stars form a single sequence? 
Why are there C-rich and N-rich stars? 
Why is there such a large range in hydrogen abundance?
How are they related to other classes of evolved star, including
normal sdB stars?
Is there a connection with any of the extreme helium
stars \citep{jeffery08b,jeffery08a}? 

Possible origins include: a late core flash of a single post-giant-branch
helium star evolving toward the white dwarf sequence
\citep{lanz04,miller08}; the merger of two helium white dwarfs
\citep{iben90,saio00}; and the merger of a helium white dwarf with a
post-sdB star \citep{justham10}. All of these scenarios are likely to
produce hot subdwarfs with He-rich and N-rich surfaces. The second has also
been argued to lead to helium-poor ``normal'' sdB stars. All predict 
evolution tracks that commence with shell helium-ignition in a white
dwarf. They take the star to a yellow-giant on a thermal time-scale and then to the helium 
main-sequence on a nuclear timescale. The details of the tracks differ
in respect of their initial conditions and the micro-physics
adopted. Whether carbon is exposed is not clear.

The goal is, if possible, to distinguish clearly the various types
of observed He-sdB 
(and He-sdO) stars and to connect each to one of these diverse origins.



The surface abundances of elements other than hydrogen and helium are 
therefore important indicators of previous evolution. But first it is
necessary to establish in what ranges these abundances lie, and hence
to identify whether distinct groups exist. The challenge is
that the numbers of He-sdB and He-sdO stars bright enough for fine
analysis has hitherto been small, and very few, so far, have been found
to be spectroscopically similar. The first of our studies was of
PG1544+488, which surprisingly turned out to be a short-period binary
containing two helium-rich sdB stars \citep{ahmad04}. The second  
was for JL\,87 \citep{ahmad07}, a relatively bright and only moderately 
helium-rich ($n_{\rm He}/n_{\rm H}\approx 0.4$) and carbon-rich subdwarf. 
Far-ultraviolet spectra of these, and of LB\,1766, were previously
analyzed by \citet{lanz04}, with quite different results to our own. 
\citet{ahmad07} demonstrated the importance of establishing $T_{\rm
  eff}$, $\log g$ and carbon abundances from optical spectra of He{\sc i} lines, {\it
  i.e.} relatively unblended lines with well-understood broadening
theory, {\it before} extracting abundances of subordinate species. 

We are therefore systematically  acquiring high-resolution high signal-to-noise
optical spectroscopy of He-sdB stars. \S\,2 describes the 
observations used in this paper.
These data are used to carry out detailed abundance analyses, 
making use of the latest generation of fully line-blanketed model 
atmospheres for appropriate mixtures (\S,3). \S\,4 presents the
results for our programme stars, which are discussed in terms 
of the evolution models and analyses of related objects in \S\,5.

\begin{table}
\centering
\caption{AAT/UCLES Observing Log}
\label{t_log}
\begin{tabular}{lllll}
\hline
\multicolumn{2}{l}{Star} & \multicolumn{2}{c}{$\alpha~~~(2000)~~~\delta$} & $m_{\rm V}$ \\
 UT (Start) & Seeing & $t_{\rm exp}$ (s) & S/N \\
\hline

\multicolumn{2}{l}{BPS\,CS\,22940--0009} & 20 30 20 & --59 50 50 & 13.8 \\
 2005 08 26 09:25:40 & 1.4'' & 1800 &10 \\
 2005 08 26 09:56:33 & 1.4'' & 1800 &10 \\   
 2005 08 26 10:27:29 & 1.4'' & 1800 &12 \\ 
 2005 08 26 10:58:22 & 1.4'' & 1800 &10 \\
 Mean                &     &        &21\\[2mm]

\multicolumn{2}{l}{LB\,1766} & 04 59 19 & --53 52 52 & 12.3 \\ 
 2005 08 26 17:22:30 & 2.0'' & 1800 &35\\
 2005 08 26 17:53:23 & 2.0'' & 1800 &35 \\
 2005 08 29 16:00:55 & 1.7'' & 1800 &20 \\ 
 2005 08 29 16:31:49 & 1.7'' & 1800 &20 \\
 Mean                &     &        &61\\[2mm]

\multicolumn{2}{l}{BPS\,CS\,22956--0094} & 22 16 56 & --64 31 51 &  --  \\
 2005 08 27 11:42:19 & 1.5'' & 1800 &17 \\
 2005 08 27 12:13:13 & 1.5'' & 1800 &13 \\ 
 2005 08 27 12:44:07 & 1.5'' & 1800 &20 \\
 Mean                &     &        &28\\[2mm]

\multicolumn{2}{l}{BPS\,CS\,29496--0010} & 23 34 02 & --28 51 38 & 14.7 \\
 2005 08 27 14:07:52 & 1.8'' & 1800  &10 \\
 2005 08 27 14:38:46 & 1.8'' & 1800  &10 \\
 2005 08 27 15:09:40 & 1.8'' & 1800  &10 \\   
 2005 08 27 15:40:37 & 1.5'' & 1800  &10 \\
 2005 08 27 16:21:17 & 1.5'' & 1800  &10 \\  
 2005 08 27 16:52:10 & 1.5'' & 1800  &10 \\   
 2005 08 27 17:23:04 & 1.5'' & 1800  &10 \\ 
 2005 08 27 17:54:36 & 1.5'' & 1800  &10 \\  
 Mean                &     &         &25\\[2mm]

\multicolumn{2}{l}{LB\,3229 = JL\,261} & 01 47 17 & --51 33 39 & 13.6 \\
 2005 08 27 18:29:28 & 1.5'' & 1800 & 17   \\
 2005 08 27 19:00:23 & 1.5'' & 1800 &18  \\    
 Mean                &     &        &25\\[2mm]

\multicolumn{4}{l}{SB\,21 = PHL\,645 = Ton-S\,137 = BPS\,CS\,29503--0009} \\
\multicolumn{2}{l}{} & 00 04 31 & --24 26 18 & 13.9 \\
 2005 08 28 14:46:45 & 1.2'' & 1800 &20 \\
 2005 08 28 15:17:39 & 1.2'' & 1800 &20 \\
 2005 08 28 15:48:33 & 1.2'' & 1800 &20 \\  
 2005 08 28 16:19:27 & 1.2'' & 1800 &19 \\   
 2005 08 28 16:50:21 & 1.2'' & 1800 &18 \\  
 2005 08 28 17:21:15 & 1.2'' & 1800 &18 \\   
 2005 08 28 17:52:09 & 1.2'' & 1800 &15 \\
 2005 08 28 18:24:53 & 1.4'' & 1800 &17 \\  
 Mean                &     &        &52\\[2mm]
\hline
\end{tabular}
\end{table}

\section{Observations}

Spectra of several hydrogen-deficient stars were obtained with the
University College  London Echelle Spectrograph on the Anglo-Australian
Telescope (AAT) in August 2005. These included a number of He-sdB
stars, namely LB\,1766, LB\,3229, SB\,21 (= Ton-S\,137 = BPS\,29503-0009),
BPS\,22940--0009, BPS\,29496--0010, and BPS\,22956--0094. An extract
from the observing log is shown in Table~\ref{t_log}, which also
indicate the signal-to-noise ratio of the combined spectra used here.  
UCLES was configured with the
31.6 lines mm$^{-1}$ grating, the EEV2 detector, and a slit width of
1.09\,mm. A central wavelength of 4340.02\,\AA\ gives complete spectral
coverage between 3820 and 5200 \AA, and a nominal resolution with this
slit width $R\approx32\,000$.
Exposures were broken into
1800\,s segments in order to minimise cosmic-ray contamination. 

A preliminary analysis of three of these stars (LB\,1766, SB\,21, and BPS
22940-0009) using the same data was given by \citep{naslim10}. 
SB\,21 was originally identified as an extremely helium-rich subdwarf \citep{hunger80}; 
\citet{hunger81} found this star to be comparable with the helium-rich hot subdwarfs, 
CPD$-31^{\circ}1701$ and TON-S\,103.  Abundances for LB\,1766 were previously obtained from a
far-ultraviolet (FUV) spectrum \citep{lanz04}; experience
has demonstrated that an analysis of the FUV spectrum alone can lead to
systematic errors \citep{ahmad07}. 
Our AAT spectra show LB\,1766 and SB\,21 to be nearly identical. 
Note that the original analyses \citep{lanz04,hunger80} suggest
these two stars to be quite different; thus a detailed
contemporary comparison is important.  

BPS\,22940--0009 and BPS\,29496--0010 were identified as He-sdB stars
and BPS\,22956--0094 as an sdB star in the survey of
\citet{beers92}. Our spectra show BPS\,22940--0009 and
BPS\,22956--0094 to be carbon-rich He-sdB stars with strong C{\sc ii},
C{\sc iii}, N{\sc ii}, N{\sc iii}, and He{\sc i} lines. In contrast, 
LB\,1766, SB\,21, BPS\,29496--0010 and LB\,3229 are carbon poor He-sdB
stars, but with strong N{\sc ii}, N{\sc iii}, and He{\sc i} lines.

The observations were
reduced using a combination of ECHOMOP routines \citep{mills96} and bespoke
\'echelle reduction software \citep{sahin08}. 
The sky-subtracted wavelength-calibrated spectrum was extracted to a 
2-d format with each order represented by a single row. 
Continuum normalization was achieved by smoothing the 2-d spectrum to
form a 2-d envelope spectrum, and then dividing by said envelope 
to remove most of the \'echelle blaze function. The smoothing
procedure was adjusted to ensure that strong lines (Balmer or He{\sc
  i} 4471 for example) were avoided when producing the normalisation function. 
Order merging was
carried out using the same 2-d envelope to provide the weights at each
wavelength in each overlap interval. A final normalisation step 
was carried out in which the merged 1-d spectrum was divided by a
low-order polynomial  fitted to a set of continuum points defined manually. 

We found no radial velocity shifts amongst the repeat spectra of
individual stars, but only the data for LB\,1766 were spread over a
significant time interval (3 days). 

In order to obtain a spectrum with the highest possible signal-to-noise 
ratio, all of the individual spectra for each object were merged together,
weighted appropriately for the signal-to-noise ratio in each
spectrum. The combined spectra were velocity-shifted to the rest frame. 

In an iterative process, we successfully identified all of the
significant absorption lines visible in the combined spectra. 
Where observations allow ({\it e.g.}
LB\,1766 and SB\,21), all known permitted and forbidden lines of He{\sc i} 
can be identified
\citep[cf. HD144941:][]{underhill66,harrison97,beauchamp98}.
He{\sc ii}4686 is present in all targets. 
Hydrogen Balmer lines are evident in BPS\,22956--0094. 
He{\sc ii}4541 and other He{\sc ii} Pickering lines are present in
LB\,3229 and BPS\,29496--0010. In the remainder, an H$\beta$/He{\sc
  ii}4859 blend is present; it is not obvious from H$\gamma$ or He{\sc
  ii}4541 which is dominant. 
In addition to N{\sc ii,iii}, the target
spectra variously show lines due to C{\sc ii,iii}, O{\sc ii}, Ne{\sc
  ii}, Mg{\sc ii}, Al{\sc iii}, Si{\sc iiii,iv} and S{\sc
  iii}. Identification charts are given in Figs.~\ref{f_LB1766} --
\ref{f_BPS29496} (online only). 

\begin{figure}
\includegraphics[angle=-90,width=0.45\textwidth]{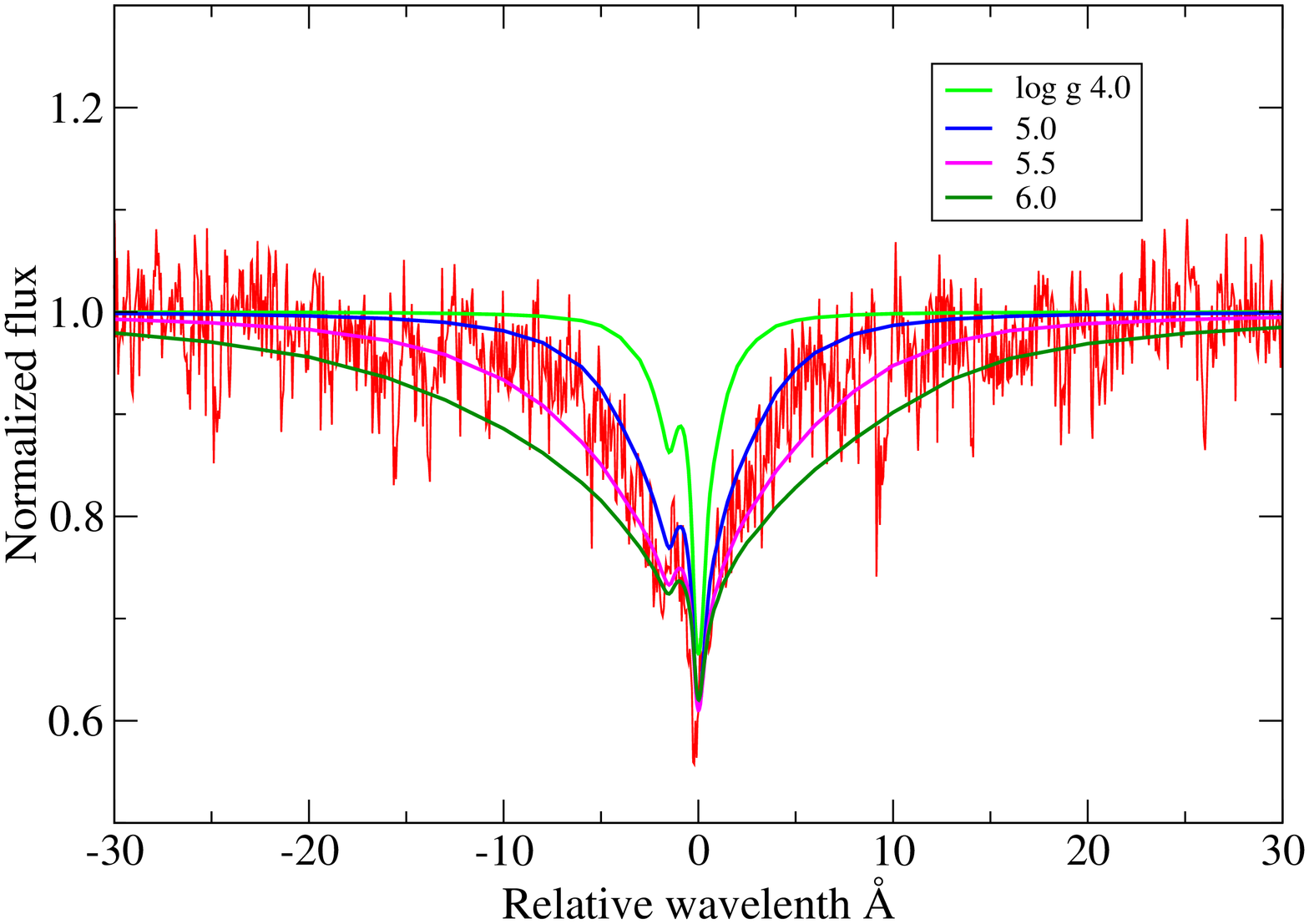}
\caption{Theoretical model fits to He{\sc i}\,4471 
in LB\,3229 for model atmospheres with $n_{\rm He}=0.989$, 
$T_{\rm eff}=40\,000\,{\rm K}$ and $\log g=4.0(0.5)6.0$.}
  \label{HeIprof}
\end{figure}

\begin{figure}
\includegraphics[angle=-90, width=0.45\textwidth]{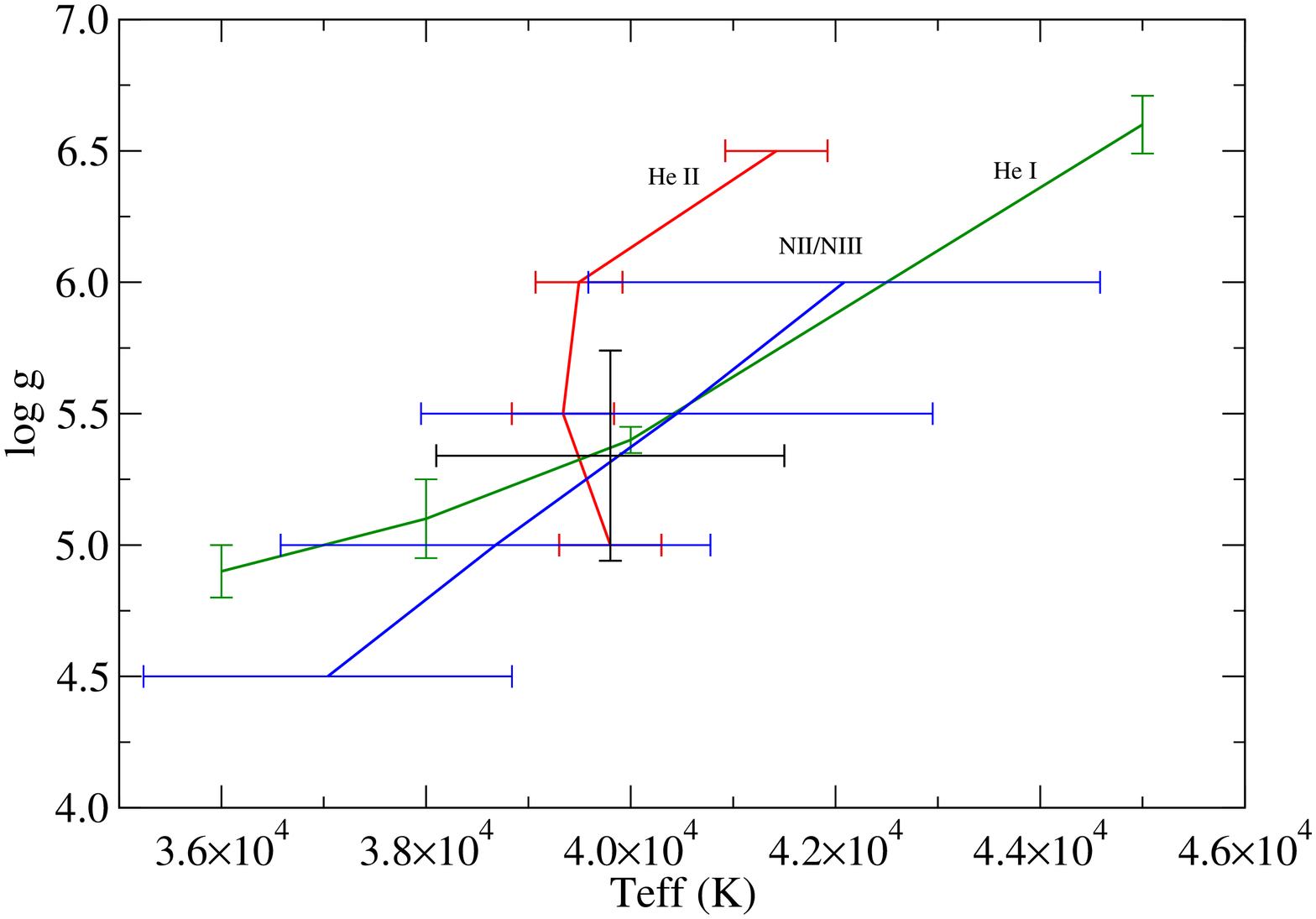}
  \caption{The loci of ionization equilibria for N{\sc ii}/N{\sc iii}
    and He{\sc ii}, the profile fits to He{\sc i} lines, and the
    adopted solution for LB\,3229.} 
  \label{ionizationeq}
\end{figure}

\section{Physical parameters of He-sdBs}

The goal was to measure atmospheric physical
parameters $T_{\rm eff}$, $\log g$ and elemental abundances for each
star. We adopted two methods to determine
$T_{\rm eff}$ and $\log g$. In the first we used ionisation
equilibria of prominent ions to determine $T_{\rm eff}$ and the profiles
of Stark-broadened He{\sc i} lines to determine $\log g$. In the
second we used the $\chi^{2}$-minimization package SFIT
\citep{ahmad03} to determine $T_{\rm eff}$ and $\log g$ simultaneously.

\subsection{Model atmospheres, line formation and spectral synthesis}

Grids of model atmospheres for hydrogen-deficient stars were
calculated using the LTE line-blanketed code STERNE \citep{behara06}
  which uses Opacity Project data for the continuous opacities, and
  treats line-blanketing through opacity sampling in a database of
  some $10^8$ atomic transitions. In this case, we adopted grids with
  1/10 solar metallicity, relative helium abundances $n_{\rm He}=
  0.50$, 0.70, 0.90, 0.95, 0.989 and 0.999 by
  number\footnote{equivalent to $n_{\rm H}=0.50$, 0.30, 0.10, 0.05,
    0.01 and 0.0, respectively.}, and assumed
  micro-turbulent velocities $v_{\rm t} = 5$ and $10\,{\rm km\,s^{-1}}$. 

The choice of the 1/10 solar metallicity grid was adopted primarily
because we were unable to obtain satisfactory fits for solar
abundance models. The average abundances of Si and Mg,
which are unlikely to have been affected by evolution, are 
sub-solar by $\approx 0.5$ dex (see below). 

Given a model atmosphere, the LTE radiative-transfer code SPECTRUM
\citep{jeffery01b} may be used to compute a) individual line profiles and
equivalent widths for given abundances and $v_{\rm t}$, b) synthetic
spectra for given wavelength ranges given the same information, and c)
the abundances of ions from individual lines of given equivalent width
and $v_{\rm t}$.

For all of the programme stars except BPS\,CS\,22956--0094, 
the Balmer lines were either weak or undetectable. 
Consequently we initially assumed a helium abundance $n_{\rm He}=0.999$
($n_{\rm H}=0.0$) and model atmospheres with 
$v_{\rm t} = 5\,{\rm km\,s^{-1}}$ as a starting approximation.  

\subsection{Micro-turbulent velocity} 

Using approximate values for $T_{\rm eff}$ and $\log g$, we measured the
micro-turbulent velocity from the equivalent widths of seventeen N{\sc
  ii} lines in LB\,1766 and SB\,21. 
Nitrogen abundances were calculated for micro-turbulent
velocities in the range $v_{\rm t}=0\,(5)\,20\,{\rm km\,s^{-1}}$ using
SPECTRUM. The micro-turbulent velocity determined by minimizing the
scatter in the nitrogen abundance was $v_{\rm t}=10\pm4\,{\rm
  km\,s^{-1}}$.  We used the measured value of $v_{\rm t}$ 
  in subsequent formal solution calculations for the ionisation 
  equilibrium and abundance measurements. 
  For the instrumental profile, we
adopted a Gaussian with FWHM = 2 resolution elements, corresponding to
0.1\,\AA\ or $R\approx45\,000$.  Additional broadening was attributed
to rotation broadening and measured as part of the SFIT solution.

\subsection{Ionisation equilibrium and He{\sc i} fitting}

Using model atmospheres with $v_{\rm t}=10\,{\rm km\,s^{-1}}$, the
ionisation equilibrium was established by balancing the abundances determined
from  N{\sc ii} and  N{\sc iii} lines in each star, 
as well as by fitting the equivalent width of He{\sc ii}\,4686. 
Values of $T_{\rm eff}$ were determined for several fixed values of $\log
g$. Ideally, all ionisation equilibria and profile fits should
converge at a single point  in parameter space. In practice
they rarely do, possibly because of departures
from LTE, which become increasingly important at $T_{\rm eff} >
30\,000$\,K especially in certain strong He{\sc i} and He{\sc ii}
lines, possibly because of errors in equivalent width measurement,
atomic data, or some other reason. 

Similarly, for a series of fixed values of $T_{\rm eff}$, we determined the
value of $\log g$ by finding the best-fit theoretical profile
for the He{\sc i} lines 4471, 4388 and 4922\,\AA\
(Fig. \ref{HeIprof}). The non-diffuse line He{\sc i}\,4121 is blended 
with nearby lines, and so was excluded from our analysis. 

The coincidence of He{\sc i} profile fits and the ionisation
equilibria was used to determine the overall solution illustrated in 
Fig.~\ref{ionizationeq}. 
Since the  He{\sc ii}\,4686 and N{\sc ii/iii} temperatures do not
coincide, and since the systematics
of these two diagnostics are not yet clear, we have chosen an
unweighted mean to determine the ionization temperature. 
The error in this mean is given by the quadratic mean of the formal
error in the  He{\sc ii}\,4686 temperature and the standard deviation in
the N{\sc ii/iii} temperature. 

In practice, this procedure requires making a choice for the helium
abundance. An estimate for $n_{\rm He}$ was obtained as described in the
following section, and the grid helium abundance closest to that 
estimate was adopted.

\begin{table*}
\centering
\caption{Atmospheric parameters}
\label{t_pars}
\begin{tabular}{@{}lllllll}
\hline
Star & $T_{\rm eff} (\rm K)$ & $\log g$ & $n_{\rm He}$ & $v_{\rm t}$& $v \sin i$ & Source\\
     &                      &          &             & $({\rm km\,s^{-1}})$ & $({\rm km\,s^{-1}})$ & \\
\hline
BPS\,CS\,22940--0009 
& $33\,700\pm800$   & $4.7\pm0.2$  & 0.993 &    & $4\pm3$ & SFIT \\
& $34\,150\pm1\,700$& $4.8\pm0.2$  &       &    &         & Ionization equilibrium \\
& $34\,000$         & $4.5$        & 0.999 & 10 &         & Adopted model \\[1mm]
BPS\,CS\,22956--0094 
& $34\,280\pm800$   & $5.63\pm0.2$ & 0.622 &    & $2\pm1$ & SFIT     \\
& $34\,100\pm2000$   & $5.52\pm0.2$ &       &    &         & Ionization equilibrium     \\
& $34\,000$         & $5.5$        & 0.699 &  5 &         & Adopted model \\[1mm]
SB 21 
& $35\,960\pm500$   & $5.4\pm0.2$  & 0.997 &    & $12\pm2$& SFIT \\
& $36\,500\pm1\,500$& $5.6\pm0.2$  &       &    &         & Ionization equilibrium \\
& $36\,000$         & $5.5$        & 0.999 & 10 &         & Adopted model \\
& $35\,000  $       & $5.4$        &       &    &         & \citet{hunger81} \\[1mm]
LB 1766 
& $36\,340\pm500  $ & $5.19\pm0.1$ & 0.997 &    & $20\pm3$& SFIT \\
& $35\,600\pm2\,100$& $5.15\pm0.25$&       &    &         & Ionization equilibrium \\
& $36\,000$         & $5.0$        & 0.999 & 10 &         & Adopted model \\
& $38\,000\pm500 $  & $5.5\pm0.3$  &       &    &         & \citet{ahmad03} \\
& $40\,000  $       & $6.3$        &       &    &         &\citet{lanz04} \\[1mm]
BPS\,CS\,29496--0010 
& $39\,150\pm1000$ & $5.65\pm0.2$ & 0.996 &    & $2\pm 1$& SFIT   \\
& $39\,770\pm2300$  & $5.8\pm0.2$  &       &    &         & Ionization equilibrium   \\
& $40\,000$         & $5.5$        & 0.999 & 10 &         & Adopted model \\[1mm]
LB\,3229         
& $40\,000\pm500$   & $5.15\pm0.2$ & 0.988 &    &$8.5\pm2$& SFIT    \\
& $39\,800\pm1700$   & $5.34\pm0.4$&       &    &         & Ionization equilibrium \\
& $40\,000$         & $5.0$        & 0.989 & 10 &         & Adopted model \\
\hline
\end{tabular}
\end{table*}

\begin{table*}
\centering
\caption{Chemical abundances }
\label{t_abs}
\begin{tabular}{@{}lcllllll}
\hline
Star & Ref & H & He & C & N & O & Ne     \\
\hline
LB\,1766 &   & $<8.5$ & 11.54 & $7.10\pm0.23$ & $8.29\pm0.25$ & $7.13\pm0.30$   & $ 8.62\pm0.42$ \\
SB\,21   &   & $<8.5$ & 11.54 & $6.73\pm0.18$ & $8.24\pm0.22$ & $7.25\pm0.27$ & $8.49\pm0.47$ \\
BPS\,CS\,29496--0010
         &   & $<8.5$ & 11.54 & $6.88\pm0.20$ & $8.48\pm0.24$ &    & \\
BPS\,CS\,22940--0009 
         &   & $9.1\pm0.2$ & 11.54 & $8.94\pm0.35$ & $8.46\pm0.22$ & $ 7.11\pm0.34$   &$8.27\pm0.45$ \\
LB\,3229 &   & $9.3\pm0.2$ & 11.54 & $7.51\pm0.27$ & $8.66\pm0.28$ & $$   &$8.62\pm0.38 $\\
BPS\,CS\,22956--0094
         &   & $11.1\pm0.2$& 11.40 & $8.52\pm0.35$ & $8.33\pm0.24$ &  $           $ &\\[2mm]

LB1766   & 1 & $9.53$      & 11.53 & $7.06$        & $8.77$        & &  \\
BX\,Cir  & 2 & 8.1         & 11.5  & 9.02 & 8.26  &  8.04 & \\
V652\,Her& 3 & 9.38        & 11.54 & 7.14 & 8.93  &  7.54 & 8.38 \\
JL\,87   & 4 & 11.62       & 11.26 & 8.83 & 8.77  &  8.60 & 8.31 \\
LS\,IV$-14^{\circ}116$
         & 5 &  11.95      & 11.23 & 8.47 & 8.23  &  \\[2mm]
Sun      & 6,7 & 12.00     &[10.93]& 8.52 & 7.92  & 8.83   & [8.08] \\[2mm]
\hline
Star & Ref & Mg & Al & Si & S \\
\hline
LB\,1766 &   & $7.17\pm0.09$  & $6.20\pm0.18$& $7.03\pm0.18$ & $6.71\pm0.20$ \\
SB\,21   &   & $ 7.24\pm0.10$ & $6.22\pm0.10$& $6.98\pm0.26$ &  $6.54\pm0.13$  \\
BPS\,CS\,29496--0010
         &   & $7.80\pm0.19$  &              & $7.05\pm0.23$ &   $6.65\pm0.15$ \\
BPS\,CS\,22940--0009 
         &   & $7.27\pm0.18$  & $6.12\pm0.15$& $7.23\pm0.24$ &   $6.45\pm0.15$ \\
LB\,3229 &   & $8.21\pm0.15$  &              & $7.38\pm0.22$ &                 \\
BPS\,CS\,22956--0094
         &   & $7.22\pm0.13$  &              & $6.82\pm0.14$ &   $6.96\pm0.10$ \\[2mm]

BX\,Cir  & 2 & 7.17     & 6.04 & 6.91 & 6.67 \\
V652\,Her& 3 & 7.76     & 6.49 & 7.49 & 7.44 \\
JL\,87   & 4 & 7.36     & 6.28 & 7.22 & 6.88 \\
LS\,IV$-14^{\circ}116$
         & 5 &  6.95    &      & 6.63 &      \\[2mm]
Sun      & 6 &  7.58    & 6.47 & 7.55 & 7.33 \\[2mm]
\hline
\end{tabular}\\
\parbox{140mm}{
References: 
1. \citet{lanz04}, 
2. \citet{drilling98}
3. \citet{jeffery99} , 
4. \citet{ahmad07},
5. \citet{naslim10c}, 
6. \citet{grevesse98}, 
7. \citet{dziembowski98} [the solar helium abundance is the
asteroseismic value for the outer convection zone, the solar neon
abundance is the meteoritic value; other solar abundances are for the solar
photosphere]. 
}
\end{table*}

\subsection{$\chi^{2}$-minimization}

We measured the physical parameters $T_{\rm eff}$, $\log g$ and hydrogen
abundance $n_{\rm H}$ using the package SFIT which finds the best-fit 
solution within a grid of synthetic spectra.  
Although $n_{\rm He}$ is a parameter of the model grid, at 
high helium abundances it is actually a proxy for  $n_{\rm H}=1-\Sigma_{i>1} n_i$;  the He{\sc i} lines are insensitive to abundance 
whilst, unusually for B stars, the Balmer lines vary strongly with
abundance.  SFIT works by fitting as many parameters and regions of
spectrum simultaneously as the user chooses. Thus it 
measures $T_{\rm eff}$ from the relative strengths of helium lines, 
the ionization equilibria of all elements in the spectrum ({\it e.g.}
He{\sc i/ii}, N{\sc ii/iii}), $n_{\rm He}$ (or $n_{\rm H}$) from the strengths of
hydrogen and helium lines and $\log g$ from profiles of
Stark-broadened lines. In our analysis, spectral regions blueward of
4050\,\AA\, where the broad wings of He{\sc i} lines merge with one
another, 
have been excluded because normalisation is difficult. 

The model grid used for the analysis of each star 
was a subset of a larger model grid in which 
$T_{\rm eff}=32\,000 - 40\,000, 50,000\,{\rm K}$, 
$\log g = 4.00 - 6.00$, 
$v_{\rm t}=10\,{\rm km\,s^{-1}}$, and 
$n_{\rm He}=0.949, 0.989, 0.999$, except in the 
case of BPS\,22956--0094, where 
$n_{\rm He}=0.500, 0.699, 0.899$ was used.
The grid spacings were $\delta T_{\rm eff}=2000\,{\rm K}$ and $\delta \log
g=0.5$. 
Since no grid was available for $n_{\rm He}=0.949$ with $v_{\rm
  t}=10\,{\rm km\,s^{-1}}$, we replaced this with a grid having 
$v_{\rm t}=5\,{\rm km\,s^{-1}}$. The spectral
 fitting was done iteratively.  In the initial iteration SFIT was run
 with three free parameters; $T_{\rm eff}$, $\log g$ and $n_{\rm He}$.
 The value of $n_{\rm He}$ was noted and fixed.  In the final
 iteration $T_{\rm eff}$ and $\log g$ were solved simultaneously.

\subsection{Atmospheric parameters}
The atmospheric parameters of each star measured 
separately using SFIT and ionisation equilibrium are given in
Table \ref{t_pars}. Both sets of results agree with one another 
to within the formal errors. In the case of LB\,1766, $T_{\rm eff}$ is lower than in both
previous studies, but close to that for SB\,21, as would be indicated
by the spectral similarity of these two stars. 
The model atmospheres are substantially
improved, incorporating more appropriate line blanketing than 
in any previous studies, and this does account for significant
shifts in $T_{\rm eff}$ in hydrogen-deficient atmospheres \citep{behara06}.

\subsection{Chemical abundances}
Having measured $T_{\rm eff}$, $\log g$ and  $n_{\rm He}$ for each
star using two different
methods, we chose the grid model atmosphere closest to these measured
values (labelled ``Model'' in  Table~\ref{t_pars}). 
We measured equivalent widths of all C, N, O, Ne, Mg, Al, Si, and S
lines for which we had atomic data. SPECTRUM can compute a curve of
growth for any given spectral line; given an equivalent width it will
then return the elemental abundances for that line.  Table
\ref{t_lines} gives the adopted oscillator strengths ($gf$), measured equivalent
widths, and line abundances. Abundances are given in the form 
$\epsilon_i = \log n_i + c$ where $\log \Sigma_i a_i n_i = \log \Sigma_i
a_i n_{i\odot} = 12.15$ and $a_i$ are atomic weights. This form 
conserves values of $\epsilon_i$ for elements whose abundances do not change, 
even when the mean atomic mass of the mixture changes substantially. 

Mean abundances for each element are 
reported in Table \ref{t_abs}; in general, the errors represent the 
standard deviation of the line abundances about the mean. However, for 
Mg, Al, and S, the error also includes the error on the equivalent width
measurement estimated from the continuum noise. 
The errors in abundance $\delta \epsilon_i$ due to a representative systematic 
change in $\delta T_{\rm eff}$ or $\delta \log g$ are shown, for three
stars, in Table~\ref{t_err}. 

The hydrogen abundances adopted previously were further refined by
starting with a model spectrum for each star defined by the best 
model (Table~\ref{t_pars}) and abundances (Table \ref{t_abs}), and by
using SFIT to solve for the hydrogen abundance by fitting the Balmer
lines only. In practice, only an upper limit of $n_{\rm H}<0.001$
could be established for three stars, while the remaining three have the abundances 
shown in Table \ref{t_abs}.

Table \ref{t_abs} compares the elemental abundances thus derived 
with those previously published for LB\,1766, with those of more H-rich ``He-sdB'' stars JL\,87 and
LS\,IV$-14^{\circ}116$, two extreme helium stars V652\,Her and BX\,Cir, and the Sun. 
These will be discussed in the next section. 

Theoretical spectra computed using the adopted grid atmosphere ``Model'' (Table \ref{t_pars}) 
and the adopted mean abundances  (Table \ref{t_abs}) are shown over-plotted on the observed spectra in 
Figs.~\ref{f_LB1766} -- \ref{f_BPS29496}. 

We note the following:\\
1) Using silicon (five to eight lines), magnesium (one line),
aluminium (two lines) and sulphur (two - three lines) as 
proxies for overall metallicity, the group is metal poor by
$\approx 0.5\pm0.2 $ dex compared with the Sun. We have not yet identified
any iron lines  in the optical spectra or analysed the FUSE spectrum of LB\,1766. \\
2) The majority are hydrogen-deficient. The Balmer lines may be
blended with weak He{\sc ii} lines. The hydrogen abundances are
measured by $\chi^2$ minimisation in the model grid; errors are
estimated. The exception is BPS\,CS\,22956--0094. \\
3) All stars are nitrogen-rich ($+0.46\pm0.11$ dex) compared with the Sun, 
and significantly so ($\approx +1$ dex) after allowing for their low metallicity. \\
4) After correcting for metallicity, the group ranges from
very carbon-rich (BPS\,22940--0009 and BPS\,22956--0094: $\approx
+0.8$ dex) to very carbon-poor (SB\,21 and BPS\,CS\,22496--0010: $\approx
-1.2$ dex). \\
5) In the two C-rich stars, the broad absoprtion at 4618\AA\ is not
satisfactorily reproduced in the model. This anomalous line was first identified
in H-deficient spectra by \citet{klemola61}, and discussed most
recently by \citet{leuenhagen96}. \\
6) In all stars (four out of six) where it can be measured, 
neon appears to be significantly overabundant. \\
7) The He{\sc i} line fits are not uniformly satisfactory. In many
cases, the observed line cores are substantially stronger than in the
theoretical profiles. Here we suspect possible non-LTE effects.  
In a few cases, particularly where the S/N ratio is low, 
one or both of the observed line wings lies below the theoretical
profile.  Here we suspect difficulties with the normalisation --
which has always been as conservative as possible.

\begin{figure}
\includegraphics[angle=90,height=8cm,width=8cm]{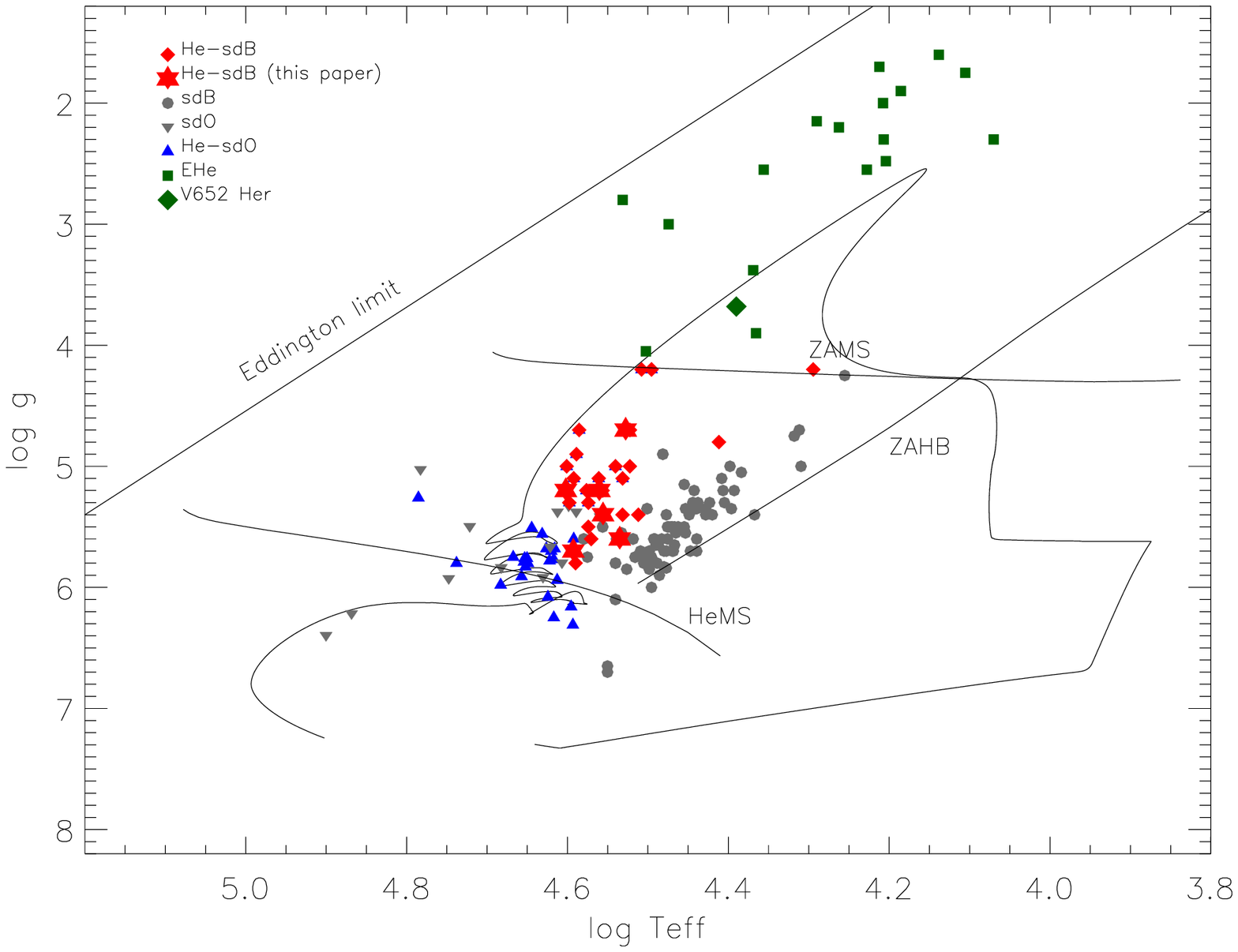}
\caption{Comparison of hot helium-rich subdwarfs and extreme helium
  stars with the evolution of a helium white dwarf with $Z=0.001$ following a late 
  helium core flash \citep{miller08}. The locations of normal sdB's 
  \citep{edelmann03} He-sdO's
  \citep{stroeer07}, He-sdB's \citep{ahmad03}, EHe's \citep{pandey06}
  and the stars discussed in this paper (N He-sdB) are identified
  separately (see key). For the latter, we adopt the SFIT results
  cited in Table 3. } 
\label{hotflasher}
\end{figure}

\begin{figure}
\includegraphics[angle=90,height=8cm,width=8cm]{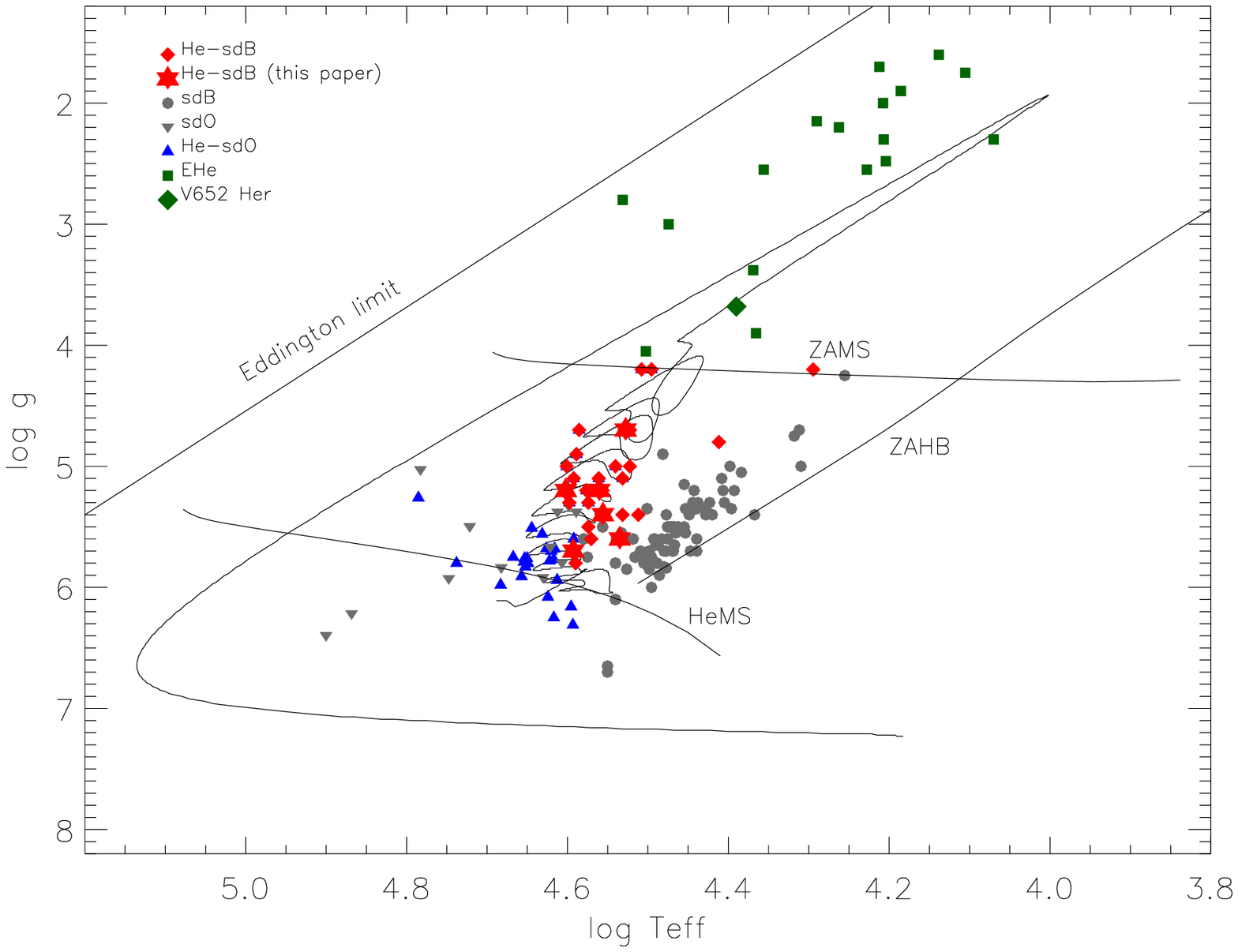}
\caption{Comparison of hot helium-rich subdwarfs and extreme helium
  stars with the evolution of a $M_{i}=0.3\,{\rm M_{\odot}}$ 
  helium white dwarf following the accretion of $0.2\,{\rm M_{\odot}}$ 
  helium to represent a double He-WD merger \citep{saio00}. Other
  symbols as for Fig.~\ref{hotflasher}
  }
\label{merger}
\end{figure}

\section{Evolutionary status of He-sdBs}

The evolutionary status of He-sdB and He-sdO stars has been discussed for some three decades. 
The location of these stars on and above the helium main-sequence points to low-mass helium stars
which are either currently in a helium-core burning phase (the helium
main sequence), or are approaching or leaving such a phase. 

There has been a tendency to analyze and interpret He-sdO 
\citep{stroeer07,napiwotzki08} and He-sdB stars \citep{lanz04,ahmad06} 
separately. This is partly due to the increasing importance of NLTE
effects in sdO stars, and hence the adoption of different model
atmospheres; a more self-consistent treatment of the two groups would
be valuable. The apparent separation of the two groups in $T_{\rm
  eff} - \log g$ space (Fig.~\ref{hotflasher}) 
may simply be a consequence of a significant range of  $T_{\rm
  eff}$ within a group of stars having comparable luminosities. 
Nevertheless, it is clear that extremely helium-rich subdwarfs 
mark out a clear locus in this diagram, and that they are more
luminous than the extended horizontal branch where hydrogen-rich sdB
stars are generally found. In addition, there are a small number of
cooler helium-rich stars, including V652\,Her and BX\,Cir which 
may be closely linked to the helium-rich subdwarfs (He-sd's). 
                                   
The question of binarity is unresolved. The prototype He-sdB star
PG1544+488 is a binary consisting of two He-sdB stars
\citep{ahmad04b}. Only one other binary
He-sd, the double He-sdO star HE\,0301--3039
\citep{lisker04} has been reported.
In such cases the extreme surface composition is a
clear consequence of a close binary interaction, probably following a
common-envelope phase, in which the entire hydrogen-rich envelope has
been ejected. 
The question is then what differentiates the production
of an He-rich subdwarf from a conventional H-rich sdB star in a close
binary.  \citet{justham10} propose a possible evolution following a
double-core common-envelope phase involving intermediate mass stars. 

In the case of a close binary where envelope ejection exposes
CNO-processed helium, the enhancement of nitrogen 
is easily explained by the conversion, and hence depletion, 
of carbon and oxygen in the CNO-cycle. A similar abundance
pattern is exhibited by the low-luminosity EHe star V652\,Her
\citep{jeffery99}.

N-rich He-sd's are less easy to explain in the absence  
of a binary companion. Equally problematic is C-enrichment in either the single-
or binary-star cases, since it requires the addition of carbon 
from $3\alpha$-burning. The low-$L$ EHe star 
BX\,Cir \citep{drilling98} provides a C-rich analogy to V652\,Her.
Neon is normally produced by $\alpha$-captures onto $^{14}$N. Since
both are plentiful in CNO-processed helium, a high-temperature episode
in the formation of the He-sd might naturally give rise to an overabundance
of neon. 

Three evolutionary models have been proposed to address these questions
for single He-sd's.

\subsection{The late hot flasher}

\citet{brown01} proposed that single star evolution with
enhanced mass loss close to the tip of the RGB will produce a star
that suffers its helium-core flash late on the white dwarf cooling
track. Since the flash occurs off-center and when the outer layers are
compact, flash-driven convection leads to mixing of the remnant
H-envelope with the helium core, and possibly also with some carbon from the
He-flash itself. The star initially expands to become a yellow giant,
and then contracts towards the He main sequence as the helium-burning
layers migrates to the center of the star. Subsequent calculations have
examined a number of variants of this model \citep{miller08} (Fig.~\ref{hotflasher}).
The result is either a N-rich or a C-rich He-sdB.

\subsection{The double helium white dwarf merger}

The merger of two helium white dwarfs has been
proposed variously to account for both normal and helium-rich hot
subdwarfs \citep{webbink84,iben90,tutukov90,saio00}. 
The progenitors are considered to be short-period systems from which 
most of the hydrogen and angular momentum has been ejected during a 
common-envelope phase. The surviving mass of hydrogen is very
small relative to the total mixed mass ({\it i.e.} that of the
disrupted white dwarf) $m_{\rm He}/m_{\rm mixed} \gtrsim
1.4\times10^{-4}/0.296$ \citet{iben86a}, 
where the hydrogen mass fraction would probably increase for lower
mass white dwarfs. 

\citet{saio00} investigated the evolution of a helium white dwarf
which rapidly accretes helium, {\it i.e.} as a result of merger with
another helium white dwarf. Following off-center helium ignition, the
star expands to become a yellow giant, and then contracts as the helium
shell burns inward through a series of mild flashes. The stable end-state of
such an evolution would most probably be an He-sdB or He-sdO star.
The surface layers of such a star should be dominated by the
nitrogen-enriched helium from the disrupted helium white dwarf. 
\citet{saio00} found no evidence for surface carbon enrichment since each
shell flash produces very little carbon and the subsequent 
flash-driven convection reaches the surface only after the 
first He-shell flash. 
The evolutionary track from one such model is shown in 
Fig.~\ref{merger}. \citet{saio00} found that one such model
would successfully account for the observed properties of the 
pulsating EHe star V652\,Her, which must be in the shell-flashing 
phase.

\subsection{The helium white dwarf plus hot subdwarf merger}

\citet{justham10} have proposed a model in
which a close binary containing a post-sdB star and a helium white
dwarf merge. The post-sdB star is essentially a $0.46\,{\rm M_{\odot}}$
hybrid white dwarf containing a small ($\approx0.3\,{\rm M_{\odot}}$) 
carbon-oxygen core and a helium envelope. The addition of fresh helium 
reignites the helium shell, and returns the star close to the helium
main-sequence with a helium-rich surface. Population synthesis
calculations yield a $T_{\rm eff} - \log g$ distribution similar to
that of the He-sdO stars of \citet{stroeer07}.

\subsection{Evolutionary status of He-sdB stars}

Comparison of the locus of He-sdB stars analysed here and by 
\citet{ahmad03} with the evolutionary calculations discussed above
(Figs.~\ref{hotflasher} and \ref{merger})
suggests that only the double helium white dwarf merger model
successfully accounts for the distribution and surface composition of
the N-rich He-sdBs. However, the evolution of both late hot
flashers and white dwarf mergers is affected strongly by the 
metallicity, mass and envelope hydrogen content. Further exploration
of the parameter space and of population statistics will be necessary
to explain the origins of these evolved stars.

\begin{figure}
\includegraphics[angle=90,width=8cm]{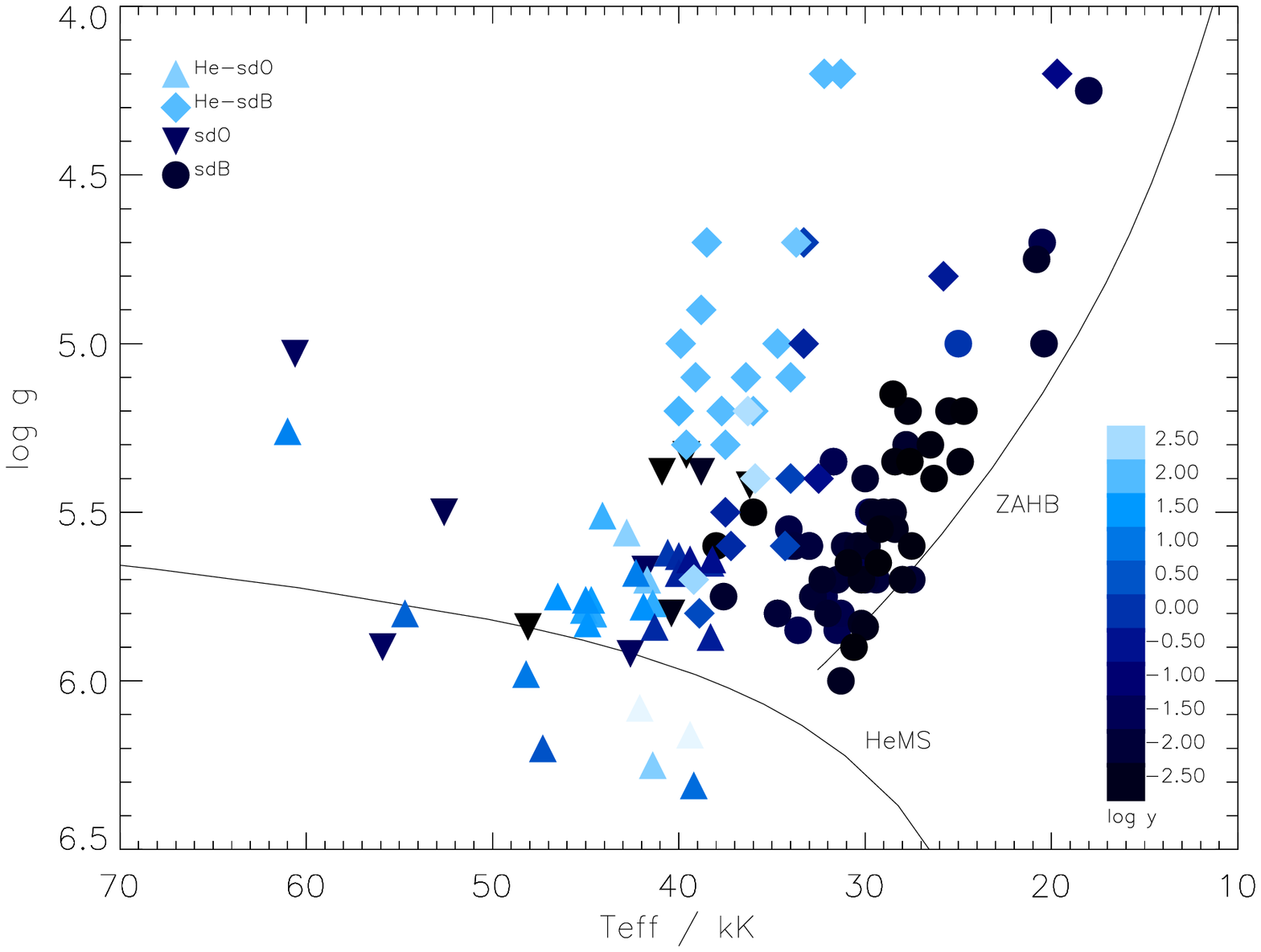}
\caption{Distribution of helium abundance versus 
$T_{\rm eff}$ and $\log g$. The helium abundance 
$\log y = \log (n_{\rm He} / n_{\rm H})$ is indicated by colour 
(or grey-scale) as shown  in the key. The data
shown include the He-sdB, He-sdO, sdB and sdO atars from
Figs.~\ref{hotflasher} and \ref{merger}.
}
\label{f_hedist}
\end{figure}



\subsection{Surface-chemistry evolution in He-sdB stars}

In all of the above evolutionary models, the initial surface composition
might be assumed to be determined by the mean mass fractions obtained
by combining several layers of stellar material. 

For example, the
surface of the double helium white-dwarf merger would comprise any
surviving hydrogen on the progenitor white dwarfs mixed with the
helium core of the less massive component. The surface hydrogen to
helium ratio (by mass) would then correspond roughly to the mass ratio of
the hydrogen and helium layers in the latter, {\it i.e.} $\approx
5\times10^{-4}$ (see above, $\equiv 2\times10^{-3}$ by number). 
Meanwhile the CNO abundances would lie somewhere between a fully CNO-cycled mixture
({\it i.e.} carbon and oxygen converted to nitrogen) and a primordial
mixture ({\it i.e.} carbon, nitrogen and oxygen scaled to the iron
abundance), depending on the helium to hydrogen fraction. Additional
carbon might be present if 3$\alpha$ burning occurs during the merger.

The surface of a late-flasher will also be represented by
CNO-processed helium, doped by whatever envelope hydrogen remained on
the surface of the giant before hydrogen-burning was
extinguished. This may be anywhere in the ranges 
$n_{\rm H}/n_{\rm He}\approx 0.001 - 0.01$ (shallow mixing) 
or $10^{-5} - 10^{-6}$ (deep mixing) \citep{miller08}. 
Carbon-enrichment may occur if flash-driven convection
can drive material to the surfce after $3\alpha$ ignition, and if the
helium-envelope mass is small compared with the amount of carbon
available for mixing \citep{cassisi03,lanz04}.

The question is then what happens to this mixture as the star
contracts towards the helium main sequence. It is well known 
that in a radiative stellar atmosphere having a sufficiently 
high surface gravity, an imbalance between the relative 
radiative and gravitational forces on individual ions leads
to chemical separation of atomic species by the process of
diffusion. Diffusion in subdwarf B stars, with $\log g>5.5$ and 
$25\,000 \lesssim T_{\rm eff}/{\rm K} \lesssim 40\,000$ \citep{heber92} 
causes hydrogen to float and helium to sink so that observed helium
abundances are typically
in the range $0.001 < n_{\rm He}/n_{\rm H}<0.01$. 

At least two factors moderate the instantaneous conversion of a contracting
He-sd into a normal H-rich sdB star. The first is that the diffusion timescale is
relatively long $\approx 10^5$y \citep[][no stellar
  wind]{unglaub01}. The evolution timescale essentially the
thermal timescale for the envelope and is $\approx 10^6$y for a merger
involving a $0.2 {\rm M_{\odot}}$ white dwarf \citep{saio00}, or $\approx
10^4$y for a late flasher with a $0.02 {\rm M_{\odot}}$ remnant
envelope \citep{miller08}.

The second factor is that a stellar wind acts to slow the diffusion
process to give a timescale  $\approx 10^6$y \citep{unglaub05,unglaub08}. Since winds in hot subdwarfs 
are radiatively driven, they are luminosity sensitive. Hence diffusion 
becomes more effective at low luminosity (high gravity). Helium
depletion in the photosphere will accelerate as a star contracts
towards the zero-age horizontal branch. 

In the absence of winds, and if all He-sd's were formed in the same
way with identical envelope masses, one might expect to see a helium
abundance gradient along the observed sequence, or at least 
to see the helium abundance drop as a helium-rich subdwarf
contracts across the ``wind-line'' -- some critical luminosity below 
which diffusion becomes effective at reducing photospheric helium. 
The fact that the most hydrogen-rich ``He-sds'' 
({\it i.e.} BPS\,CS\,22956-0094, JL\,87, LS\,IV$-14^{\circ}116$)
 are also those closest to the sdB domain might be important in this regard. 

To explore such an hypothesis, Fig.~\ref{f_hedist} shows the distribution of hot
subdwarfs as a function of helium  abundance $\log y = \log (n_{\rm
  He} / n_{\rm H})$;  the sample is
exclusively that given in Figs.~\ref{hotflasher} and \ref{merger}. 
These observations indicate a substantial majority of normal sdB stars with
negligible helium, a significant  number of extremely He-rich stars on the pre-subdwarf
cooling track and 
a continuum of hot subdwarfs with relatively high gravities and 
intermediate helium abundances ($-1 < \log y < +1$). 
In numerical terms, there are roughly 100 normal sdBs for
every ten helium-rich plus intermediate-helium subdwarfs \citep{green86}, and
eighteen helium-rich He-sdBs for nine intermediate He-sdBs
(Fig.~\ref{f_hedist}); better statistics would be valuable. 

Hot subdwarfs with intermediate helium abundances might then represent
stars in which diffusion has started to operate, but in which some
helium remains visible. The ratio of intermediate helium subdwarfs to
normal subdwarfs should then be given approximately by the ratio of the diffusion
timescale ($\approx10^6$ y) to the sdB nuclear timescale
($\approx10^8$ y) times the fraction  of sdB stars formed through
channels which involve a helium-rich progenitor ($\lesssim0.5$),
leading to a total of roughly one intermediate helium subdwarf in
one hundred, slightly fewer than observed.

One problem with this argument, and there are many, is
that helium-rich sdB stars may simply become helium-rich sdO stars on
the helium main sequence and may {\it not} evolve into helium-poor sdB
stars. A more detailed examination of surface abundances of other
species, including iron, in several helium-rich hot subdwarfs will be necessary.

One observation, however, is instructive. Amongst the helium-rich
subdwarfs studied by \citet{ahmad06}, \citet{stroeer07} and ourselves, stars
with intermediate helium abundances lie predominantly at the boundary
between the He-poor and He-rich subdwarfs in the $\log g-T_{\rm eff}$
diagram. There are virtually no He-poor
subdwarfs significantly above the horizontal branch
(Fig.~\ref{f_hedist}), 
{\it i.e.} with $T_{\rm eff} > 30\,000 {\rm K}$ and $\log g < 5.3$.
This observation is supported
by low-resolution classification surveys \citep{winter06,drilling03}. 
{\it Thus, if subdwarfs evolve onto either
the helium main sequence or the extended horizontal branch by
contracting from a more expanded configuration, then the only ones
which are currently observed to be doing so are helium rich. }

\section{Conclusion}
As part of an extended study of the surface abundances of extremely 
helium-rich hot subdwarfs, high-resolution optical \'echelle spectra 
of the He-sdB stars LB\,1766, SB\,21,  BPS\,CS\,22940--0009, BPS\,CS\,29496--001,
BPS\,CS\,22956--0094 and LB\,3229 have been presented. Opacity-sampled line-blanketed
model atmospheres have been used to derive atmospheric properties and 
surface abundances. 

All the stars analysed are moderately metal poor compared
with the Sun ([Fe/H]$\approx$--0.5). LB\,1766 and SB\,21, BPS\,CS\,29496--001 and
LB\,3229 are nitrogen-rich He-sdBs, while  BPS\,CS\,22940--0009 and BPS\,CS\,22956--0094 are
carbon-rich He-sdBs. The former have a surface composition and $L/M$
ratio comparable with the extreme helium star V652\,Her, 
while the latter might be more directly compared with the extreme
helium star BX\,Cir. 

The evolutionary status of He-sdB's has been discussed in the context
of i) close-binary star evolution, ii) a late helium flash in a post-RGB
star, iii) the  merger of two helium white dwarfs, and iv) the merger of a
helium white dwarf with a post-sdB star. The surface composition and 
locus of single N-rich He-sdBs are currently best explained by the 
merger of two helium white dwarfs, although this may not necessarily
be an unique solution. The merger of a helium white dwarf with a
post-sdB white dwarf offers an interesting alternative. 
C-rich He-sdBs require further investigation;
the origin of surface carbon is difficult to explain without mixing
$3\alpha$ products to the surface and so far, only the late 
flasher model seems capable of this. An over-abundance of neon
requires further explanation. 

On the basis of any of these evolution tracks, the EHe stars 
V652\,Her and BX\,Cir are likely to evolve to become He-sdB stars. 
He-sdB stars are likely to evolve to become He-sdO stars.

\section*{Acknowledgments}
This paper is based on observations obtained at the Anglo-Australian
Telescope. 
It has made use of the SIMBAD database,
    operated at CDS, Strasbourg, France.
The Armagh Observatory is funded by direct grant from the Northern
Ireland Dept of Culture Arts and Leisure.

\bibliographystyle{mn2e}
\bibliography{mnemonic,hesdb_abunds}

\label{lastpage}


\clearpage

\appendix
\section{Online Material}

\newpage

\begin{table*}
\centering
\caption{Oscillator strength ($\log gf$), measured equivalent width
  ($w_{\lambda}$) and derived elemental abundance $\epsilon_i$ for each line
  measured in the six programme stars.}
\label{t_lines}
\begin{tabular}{@{}cccccccccccccc}
\hline

\multicolumn{1}{c}{Ion} &
\multicolumn{1}{c}{} &
\multicolumn{2}{c}{LB\,1766} &
\multicolumn{2}{c}{SB\,21} &
\multicolumn{2}{c}{BPS\,22940--0009}&
\multicolumn{2}{c}{BPS\,29496--0010}&
\multicolumn{2}{c}{ BPS\,22956--0094}&
\multicolumn{2}{c}{LB\,3229}\\
\multicolumn{1}{c}{$\lambda({\rm \AA})$} &
\multicolumn{1}{c}{$\log gf$} &

\multicolumn{1}{c}{$w_{\lambda}({\rm m\AA})$} &
\multicolumn{1}{c}{$\epsilon_{i}$} &
\multicolumn{1}{c}{$w_{\lambda}({\rm m\AA})$} &
\multicolumn{1}{c}{$\epsilon_{i}$} &
\multicolumn{1}{c}{$w_{\lambda}({\rm m\AA})$} &
\multicolumn{1}{c}{$\epsilon_{i}$} &
\multicolumn{1}{c}{$w_{\lambda}({\rm m\AA})$} &
\multicolumn{1}{c}{$\epsilon_{i}$}&
\multicolumn{1}{c}{$w_{\lambda}({\rm m\AA})$} &
\multicolumn{1}{c}{$\epsilon_{i}$} &
\multicolumn{1}{c}{$w_{\lambda}({\rm m\AA})$} &
\multicolumn{1}{c}{$\epsilon_{i}$} \\

\hline
  C\,{\sc ii}       &      &       &       &      &        &      &         &        &    &    &  &   &   \\  
4267.02     & 0.559$\rceil$      &       &       &      &        &  490     & 9.21            &        &    & 398   &8.74  &   &   \\
4267.27     & 0.734$\rfloor$     &       &       &      &        &     &        &    &   &    &   &   \\
4074.52     & 0.408$\rceil$      &       &       &      &        & 276 & 9.13             &        &    &   123   &8.49  &   &   \\
4074.85     & 0.593$\rfloor$     &       &       &      &        &  &&        &    &  &     &   &   \\
4075.94     & -0.076$\rceil$     &       &       &      &        & 350     &  8.90          &        &    &  200    &8.47    &   &   \\
4075.85     & ~0.756$\rfloor$    &       &       &      &        &  &&        &    &   &     &   &   \\
4074.48     & 0.204              &       &       &      &        & 291     &  9.19               &        &    & 123   & 8.49 &   &   \\
4374.27     & 0.634              &       &       &      &        & 169     & 9.05             &        &    & 75   &8.52  &   &   \\[1mm]

  C\,{\sc iii} &                &       &       &      &        &         &         &        &    &    &  &   &   \\
  4647.42   & 0.072             &  100  &  7.27 & 31   & 6.61   & 380     & 9.04            & 65     & 6.97    &189    &8.35   & 90  &7.22  \\
  4650.25   & -0.149            &  52   &  6.94 & 33   & 6.86   & 262     & 8.71            & 35     & 6.80    &136    &8.16    &83   &  7.37 \\
 4651.47    & -0.625            &       &       &      &        & 256     & 9.15            &        &    &121    &8.51  &80   &7.82   \\
  4067.94   & 0.827             &       &       &      &        & 231     & 8.30                &        &    &73    &7.64  &   &   \\
  4068.91   & 0.945             &       &       &      &        & 171     & 7.94            &        &    &    &  &   &   \\
  4070.26   & 1.037             &       &       &      &        & 194     & 8.97           &        &    & 153   &8.99  &   &  \\
4186.90     & 0.924             &       &       &      &        & 252     & 9.16            &        &    &222    &9.15  &64   & 7.64   \\
4156.74     & 0.842$^1$             &       &       &      &        & 170     &  8.65                &        &    &140    &8.96  &   &   \\
4162.87     &0.218              &       &       &      &        & 140     & 8.99            &        &    &170    &9.60  &   &   \\[1mm]
 
  N\,{\sc ii}       &           &       &       &      &        &         &         &        &    &    &  &   &   \\ 
 3995.00    & 0.225     &  125  & 8.14  & 110  &  7.94  & 146     & 8.25                &189     &8.91    &124  &8.18  & 84  & 8.67  \\
 4041.31    & 0.830     &  103  & 8.07  & 134  &8.19    & 109     & 8.10                &150     &8.48    &136  &8.35  & 80  & 8.71  \\
 4043.53    & 0.714     &  76   & 7.94  & 107  &8.11    & 108     & 8.21                &81      &8.13    &85   &8.07  & 46  & 8.46  \\
 4056.90    & -0.461    &  28   & 8.51  & 31   &8.50    &         &         &        &        &     &      &     &   \\
 4073.05    & -0.160    &  53   & 8.56  & 77   &8.72    &         &         &        &        &     &      &     &   \\
 4171.59    & 0.281     &  48   & 8.09  & 70   &8.24    &  70     & 8.30                &49      &8.28    &41    &8.06  &   &   \\
 4176.16    & 0.600     &  79   & 8.09  & 85   &8.05    &  119    & 8.43                &65      &8.12    &80    &8.15  &70   & 8.86  \\
 4236.98    & 0.567     &  139  & 8.66  & 135  &8.51    &  126    & 8.55                &143     &8.74    &161    &8.82  &63   &8.83   \\
 4241.78    & 0.728     &  179  & 8.78  & 83   &7.95    &  101    & 8.17               &128     &8.49    &98    &8.21  & 105  &9.06  \\
 4447.03    & 0.238     &  59   & 7.85  & 65   &7.88    &  100    &  8.21               &94      &8.47    &118    &8.53  & 85  &8.98   \\
 4530.40    & 0.671     &  84   & 8.17  & 83     &8.08  & 122     &  8.49               &128     &8.60    &100    &8.36  &   &   \\
 4643.09     & -0.385   &  78   & 8.37  & 80     &8.35   & 117    & 8.66                &        &        &74    &8.39  &   &   \\
 4630.54    &  0.093    &  96   & 8.07  & 103    &8.11   & 139    & 8.37                &63      &8.02    &97     &8.15  &   &   \\
 4621.29    & -0.483    &  74   & 8.42  & 64     & 8.27  &        &         &        &        &73    &8.48  &   &   \\
 4613.87    & -0.607    &  71   & 8.51  & 74     &8.50   & 91     &  8.63               &57      &8.64    &82    & 8.69 &   &   \\
 4601.48    & -0.385     &   83   & 8.41  & 70   & 8.24   & 126    & 8.74                &66      &8.52    &98    &8.63      &    &   \\
4607.16     & -0.483     &   83   & 8.41  & 85   & 8.50   &122      & 8.80            &62      &8.57    &64    &8.38      &    &       \\[1mm]

  N\,{\sc iii}  &            &        &       &      &        &        &         &        &        &       &       &      &     \\
 4097.33    & -0.066     &  169   & 8.15  & 144   &8.14   & 211    & 8.42            &229     &8.51    &110    &7.97   & 187  &8.09  \\
 4103.43    & -0.377     &  133   & 8.14  & 112   &8.16   & 216    & 8.77            &225     &8.80    &       &       & 243  & 8.86  \\
 4640.64    & 0.140      &  162   & 8.40  & 104   &8.15   & 200    & 8.58            &155     &8.30    &113    &8.38   &227   & 8.51  \\
 4634.14    & -0.108     &  135   & 8.45  & 110   &8.47   & 170    & 8.64            &174     &8.70    &63     &8.11   & 186  & 8.50  \\
 4195.76$^2$ & -0.018     &  83    & 8.82  & 72    &8.88   &        &         &        &        &       &       &109   &8.74   \\
 4200.10$^2$ & 0.241      &  92    & 8.67  & 89    &8.84   &        &         &        &        &65     &8.73   &85    & 8.24  \\
4641.85     & -0.815     &        &       &       &       &        & &        &        &       &       &95    & 8.44  \\[1mm]

 O \,{\sc ii }      &            &        &       &       &       &        &         &        &        &       &       &      &   \\
    4649.14    & 0.342          &  31      & 7.08      & 44      &  7.26     &  24      & 6.87            &        &        &       &     & &  \\
 4414.90    & 0.210          &  30      & 7.25      & 38      &  7.34     &        &             &        &        &      &      &   &   \\
 4072.15    & 0.552          &  21      & 7.06      & 24      &  7.09     &  40      &  7.35           &        &        &      &      &     &       \\
 4416.97$^3$&                & 61      & 7.92      & 55      & 7.82          &          &                 &        &        &      &      &     &       \\
 4075.86    &              &          &           &    49   &    7.33 &          &                 &        &        &      &      &     & \\[1mm]

 Ne\,{\sc ii}&         &        &       &       &       &        &         &        &    &    &  &   &   \\
 4219.76    & -0.150          &  40      &  8.65     &  75     &  9.03     &  55         & 8.82             &        &        &      &      &   40   &  9.16     \\
 4231.60    & -0.450          &  39      &  8.94     &  39     &  8.95     &           &          &        &        &      &      &      &       \\
 4430.90    & -0.609          &  43      &  9.22     &  28     &  9.00     &  24      &  8.87           &        &    &    &  &      &       \\
 4397.94    & -0.160          &  35      &  8.63     &  36     &  8.66     &        &         &        &    &    &  &     &       \\
 4409.29    &  0.680         &   55     &    8.08   &    55   &   8.08    &    53       & 8.03             &        &        &      &      &  35  &  8.30     \\
 4413.20    &  0.550          &  35   &      7.94     &  34     & 7.94      &  48         &8.10              &        &        &      &      &      &       \\
 4290.37    &  0.920$\rceil$          &  100      &  8.77     &  63     & 8.13      &  45      &  7.91           &        &    &    &  &  58    & 8.50      \\
 4290.60    &  0.830$\rfloor$          &           &           &         &           &          &             &        &    &    &  &       &       \\[1mm]
\hline
\end{tabular}
\parbox{170mm}{
$gf$ values: 
C{\sc ii} \citet{yan87}, 
C{\sc iii} \citet{hib76,har70,boc55}, 
N{\sc ii} \citet{bec89}, 
N{\sc iii} \citet{but84}, 
O{\sc ii} \citet{Bec88}, 
Ne{\sc ii} \citet{wie66} \\
Notes: 1: empirical oscillator strength to match observed line; not used in mean. 
2: blended with N{\sc ii}; not used in mean, except for LB\,3229 where N{\sc ii} is weak. 
3: possibly blended; not used in mean.
}
\end{table*}

\newpage
\addtocounter{table}{-1}
\begin{table*}
\centering
\caption{contd.}
\begin{tabular}{@{}cccccccccccccc}
\hline

\multicolumn{1}{c}{Ion} &
\multicolumn{1}{c}{} &
\multicolumn{2}{c}{LB\,1766} &
\multicolumn{2}{c}{SB\,21} &
\multicolumn{2}{c}{BPS\,22940--0009}&
\multicolumn{2}{c}{BPS\,29496--0010}&
\multicolumn{2}{c}{BPS\,22956--0094}&
\multicolumn{2}{c}{LB\,3229}\\

\multicolumn{1}{c}{$\lambda({\rm \AA})$} &
\multicolumn{1}{c}{$\log gf$} &

\multicolumn{1}{c}{$w_{\lambda}({\rm m\AA})$} &
\multicolumn{1}{c}{$\epsilon_{i}$} &
\multicolumn{1}{c}{$w_{\lambda}({\rm m\AA})$} &
\multicolumn{1}{c}{$\epsilon_{i}$} &
\multicolumn{1}{c}{$w_{\lambda}({\rm m\AA})$} &
\multicolumn{1}{c}{$\epsilon_{i}$} &
\multicolumn{1}{c}{$w_{\lambda}({\rm m\AA})$} &
\multicolumn{1}{c}{$\epsilon_{i}$}&
\multicolumn{1}{c}{$w_{\lambda}({\rm m\AA})$} &
\multicolumn{1}{c}{$\epsilon_{i}$} &
\multicolumn{1}{c}{$w_{\lambda}({\rm m\AA})$} &
\multicolumn{1}{c}{$\epsilon_{i}$} \\

\hline
Mg\,{\sc ii}      &      &       &       &       &       &        &         &        &    &    &  &   &   \\
4481.13    & 0.568  &  61 & 7.17 &  77     &7.24   & 88     &  7.27   & 135    &7.80    &76  &7.22   &114 &8.21  \\[1mm]
 
Al\,{\sc iii}     &      &        &       &       &       &        &         &        &    &    &  &   &   \\
4512.54    &  0.405 &  40 & 6.34 &  34 & 6.21 &  29      &  6.10           &        &        &    &       &    &      \\
4529.20    &  0.660 &  38 & 6.05 &  54 & 6.22 &  49      &  6.14           &        &        &    &       &    &      \\[1mm]

Si\,{\sc iii}  &            &        &       &       &       &        &         &        &        &       &       &      &   \\
4552.62    &  0.283 & 114 & 6.96 & 114 & 6.88   & 164    & 7.25            &75      &6.99    & 104   &6.90 & &  \\
4567.82    &  0.061 &  82 & 6.91 &  69 & 6.71   & 138    & 7.28            &67      &7.14    & 75   &6.87  &   &   \\
4574.76    & -0.416 &  59 & 7.16 &  60 & 7.09   & 98     & 7.43            &        &        & 39   &6.94  &   &   \\
4828.96    &  0.924 &  64 & 7.07 &     &       &        &         &        &        &      &      &      &   \\
4819.72    &  0.814 &  68 & 7.22 &     &       &        &         &        &        &      &      &     &   \\
4813.30    &  0.702 &  53 & 7.18 &     &       &        &         &        &        &      &      &     &   \\[1mm]

Si\,{\sc iv}      &      &       &     &       &       &        &         &        &    &    &  &   &   \\
4088.85    &  0.199 & 143 & 6.70 & 123 & 6.72   & 237       & 7.35         &139     &6.77    &97    &6.58  &250   & 7.71  \\
4116.10    & -0.103 & 147 & 7.04 & 129 & 7.09   & 136       & 6.82     &167     &7.32    & 89   &6.80  &157   &7.26   \\
4654.14    &  1.486 &     &      & 159 & 7.41   &        &         &        &    &    &  &155   &7.32   \\
4212.41    &  0.804 &     &      &     &       &        &         &        &    &    &  &71   &7.24   \\[1mm]

S \,{\sc iii}  &    &     &      &     &      &        &         &        &    &    &  &   &   \\
4253.59    &  0.233 &  78 & 6.78 &  51 & 6.46 &  60 & 6.47 &  31 & 6.55      &  82    & 6.91     &      &       \\
4284.99    & -0.046 &  33 & 6.52 &  32 & 6.48 &  31 & 6.38 &  25 & 6.74       & 60     & 6.97     &      &       \\
4332.71    & -0.393 &  31 & 6.83 &  25 & 6.70 &  20 & 6.50 &     &
 & 36 &7.00  &      &       \\[1mm]
\hline
 
\end{tabular}
\parbox{170mm}{
$gf$ values: 
Si{\sc iii} \citet{bec90,har70}, 
Si{\sc iv} \citet{bec90,kur75}, 
Mg{\sc ii} \citet{wie66},
O{\sc ii} \citet{Bec88}, 
Al{\sc iii} \citet{top92,McE83}, 
S{\sc iii} \citet{wie69, har70}
}
\end{table*}

\begin{table*}
\centering
\caption{Abundance errors $\delta \epsilon_i$ due to errors
  in $T_{\rm eff}$ and $\log g$.}
\label{t_err}
\begin{tabular}{@{}lllllllll}
\hline
Star  & \multicolumn{8}{l}{$\delta \epsilon_i$} \\
      & C & N & O & Ne & Mg & Al & Si & S \\
\hline
BPS\,22940--0009&  &  &  & &  &  &  &     \\
$\delta T_{\rm eff}=1000\,\rm K$ &+0.03 &+0.06 &+0.11 &+ 0.05     &+0.08      &+0.10  &+0.08 &+0.14   \\
$\delta \log g=0.2$ &+0.008 &+0.004 & --0.02      &+0.01     &--0.03
&--0.02  &+0.02 &--0.02        \\[1mm] 
 
LB\,1766&  &  &  & &  &  &  &     \\
$\delta T_{\rm eff}=1000\,\rm K$ &--0.07 &+0.09 &+0.11 & +0.05     & +0.08      & +0.09  & +0.10 &+0.14        \\
$\delta \log g=0.2$ &+0.10 &+0.004 & --0.03      & +0.008     &--0.04
& --0.04  & --0.02 &--0.03        \\[1mm] 

LB\,3229&  &  &  & &  &  &  &     \\
$\delta T_{\rm eff}=1000\,\rm K$ &+0.10 &+0.16 & &  +0.17    & +0.19      &   & +0.11  &      \\
$\delta \log g=0.2$ &--0.04 &--0.09 &        & --0.10     & --0.12  &   & --0.04  &         \\
\hline
\end{tabular}\\

\end{table*}

\newpage
\begin{figure*}
\centering
\includegraphics[trim=0cm 3cm 0cm 0cm, height=9.0in,width=7.5in]{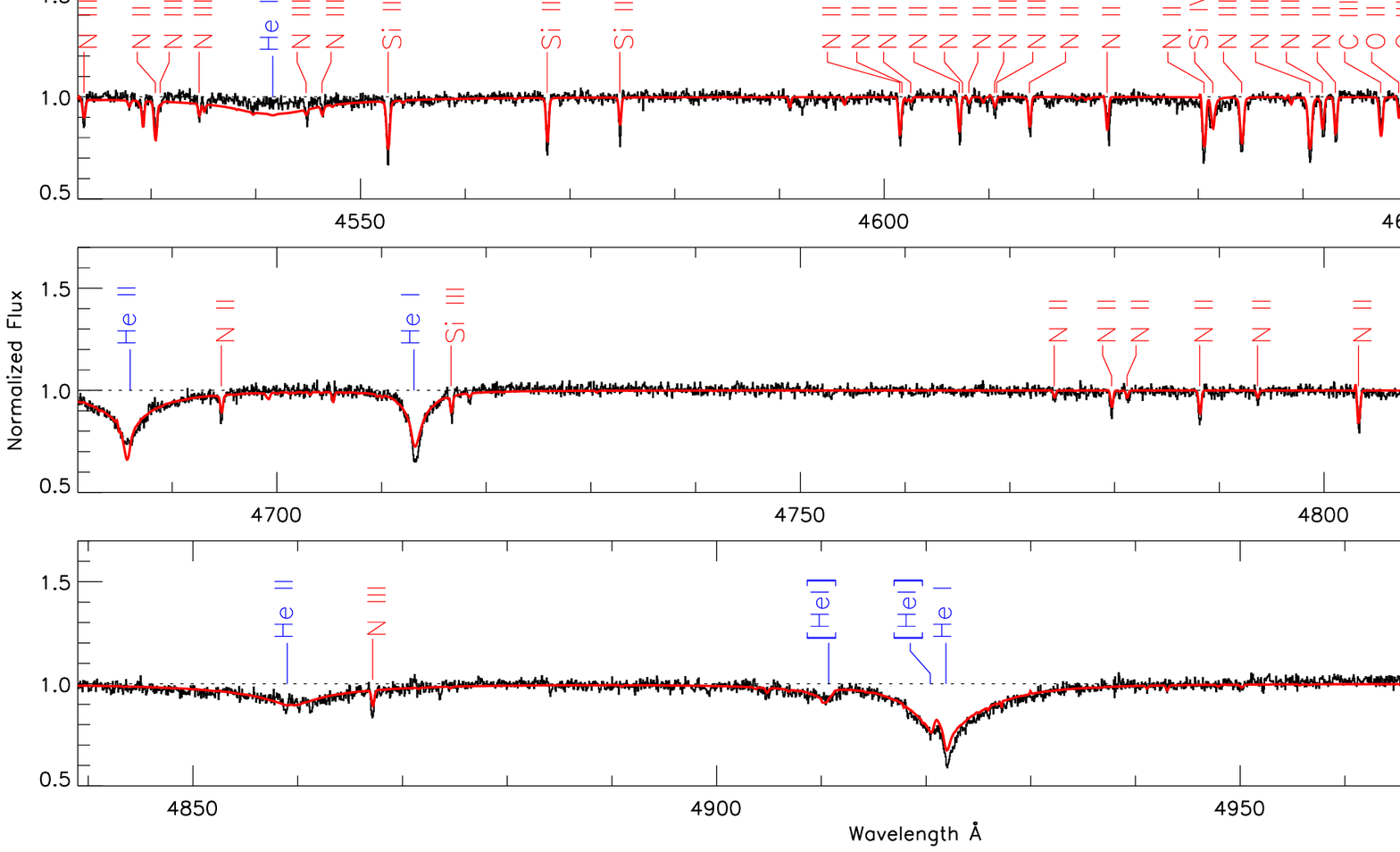}
\caption{The AAT spectrum of LB\,1766 along with the best-fit model
  of $T_{\rm eff}=36\,000\,{\rm K}$, $\log g=5.0$, $n_{\rm He}=0.999$, $v
  \sin i=20\,{\rm km\,s^{-1}}$, and $v_{\rm t}=10\,{\rm km\,s^{-1}}$. Abundances are as in Table~\ref{t_abs}. }
\label{f_LB1766}
\end{figure*}

\begin{figure*}
\centering
\includegraphics[trim=0cm 3cm 0cm 0cm, height=9.0in,width=7.5in]{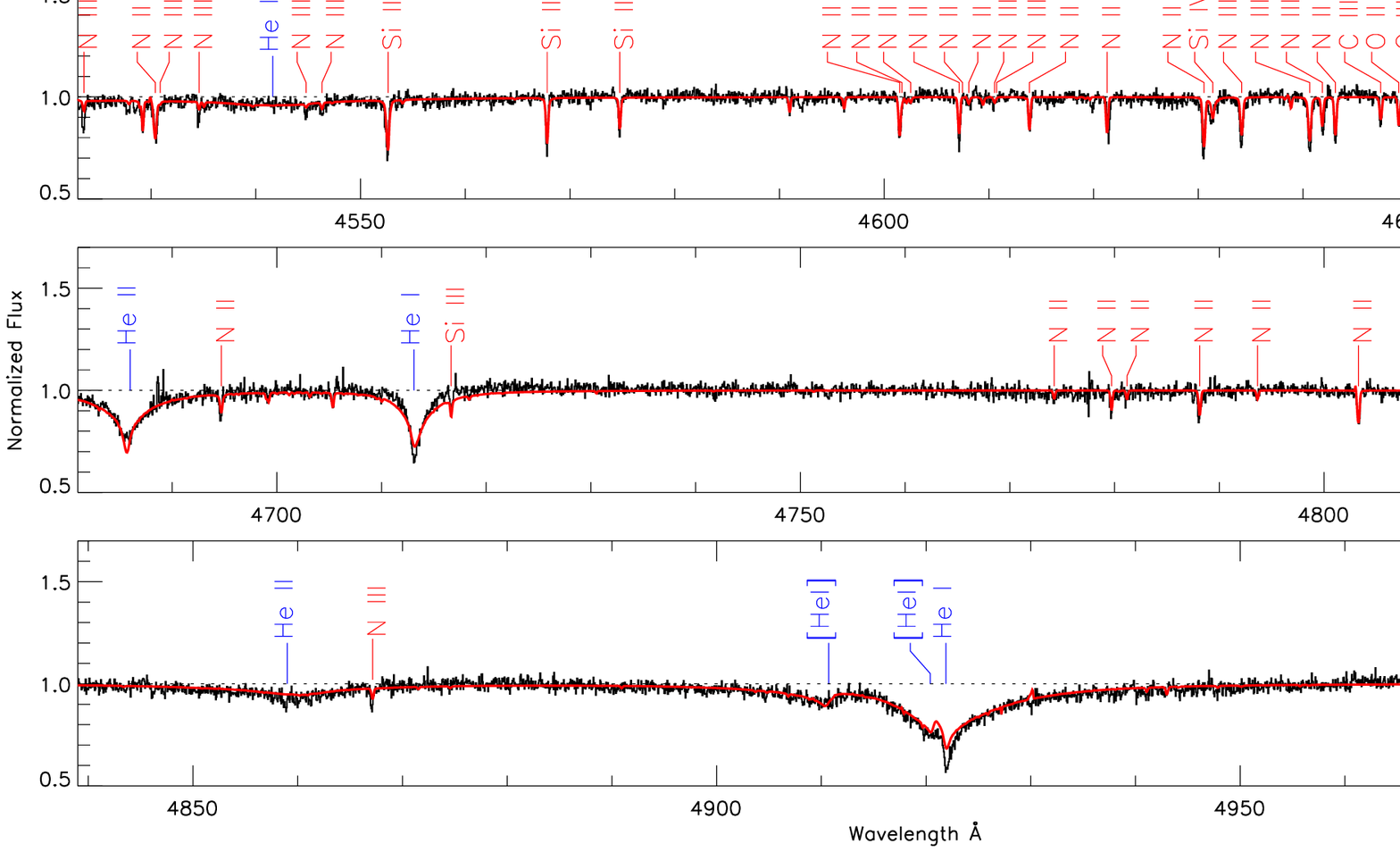}
\caption{The AAT spectrum of SB\,21 along with the best-fit model
  of $T_{\rm eff}=36\,000\,{\rm K}$, $\log g=5.5$, $n_{\rm He}=0.999$, $v
  \sin i=12\,{\rm km\,s^{-1}}$, and $v_{\rm t}=10\,{\rm km\,s^{-1}}$. Abundances are as in Table~\ref{t_abs}. }
\label{f_SB21}
\end{figure*}

\begin{figure*}
\centering
\includegraphics[trim=0cm 3cm 0cm 0cm,height=9.0in,width=7.5in]{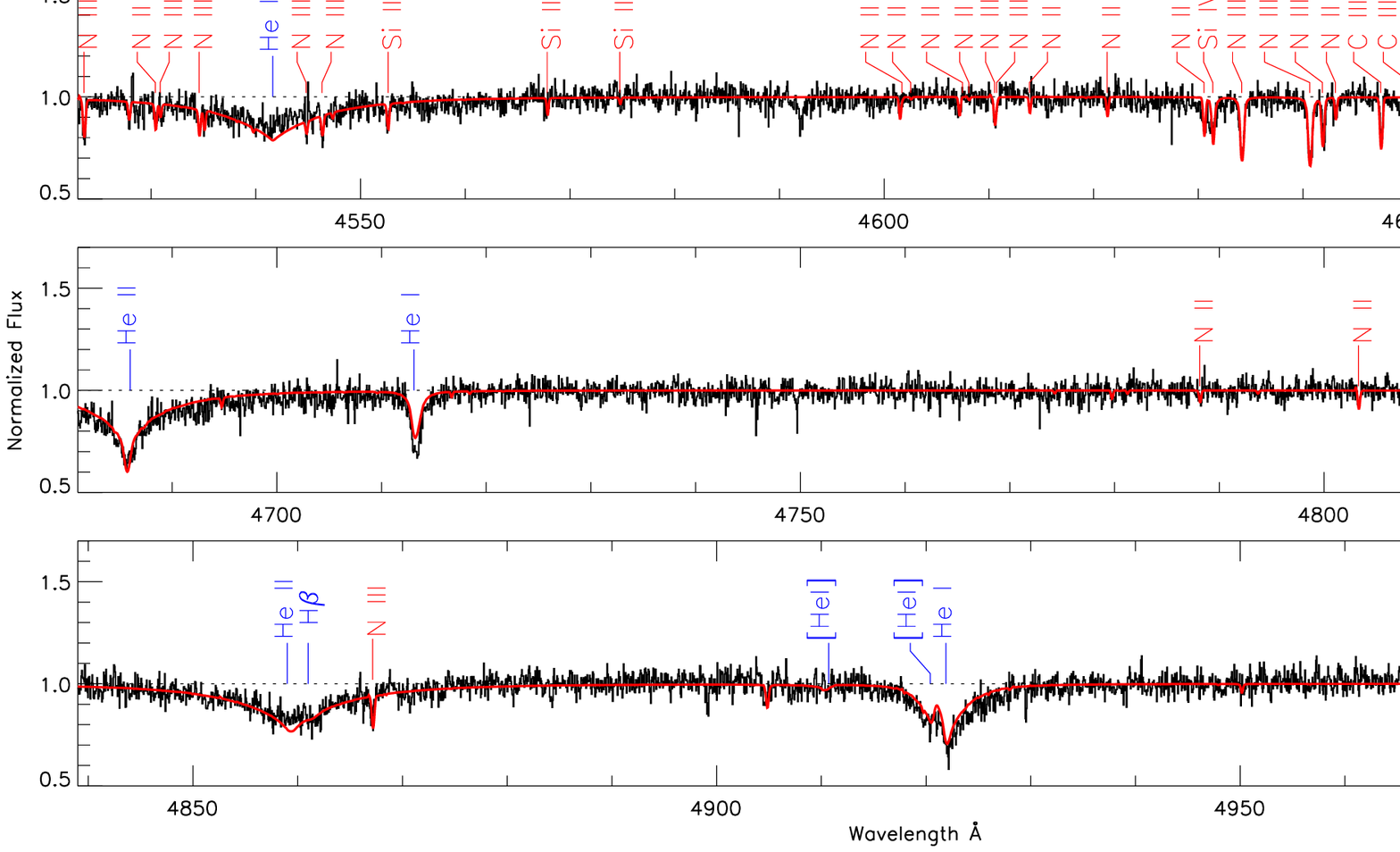}
\caption{The AAT spectrum of LB\,3229 along with the best-fit model
  of $T_{\rm eff}=40\,000\,{\rm K}$, $\log g=5.0$, $n_{\rm He}=0.989$, $v
  \sin i=8.5\,{\rm km\,s^{-1}}$, and $v_{\rm t}=10\,{\rm km\,s^{-1}}$. Abundances are as in Table~\ref{t_abs}. }
\label{f_LB3229}
\end{figure*}

\begin{figure*}
\centering
\includegraphics[trim=0cm 3cm 0cm 0cm, height=9.0in,width=7.5in]{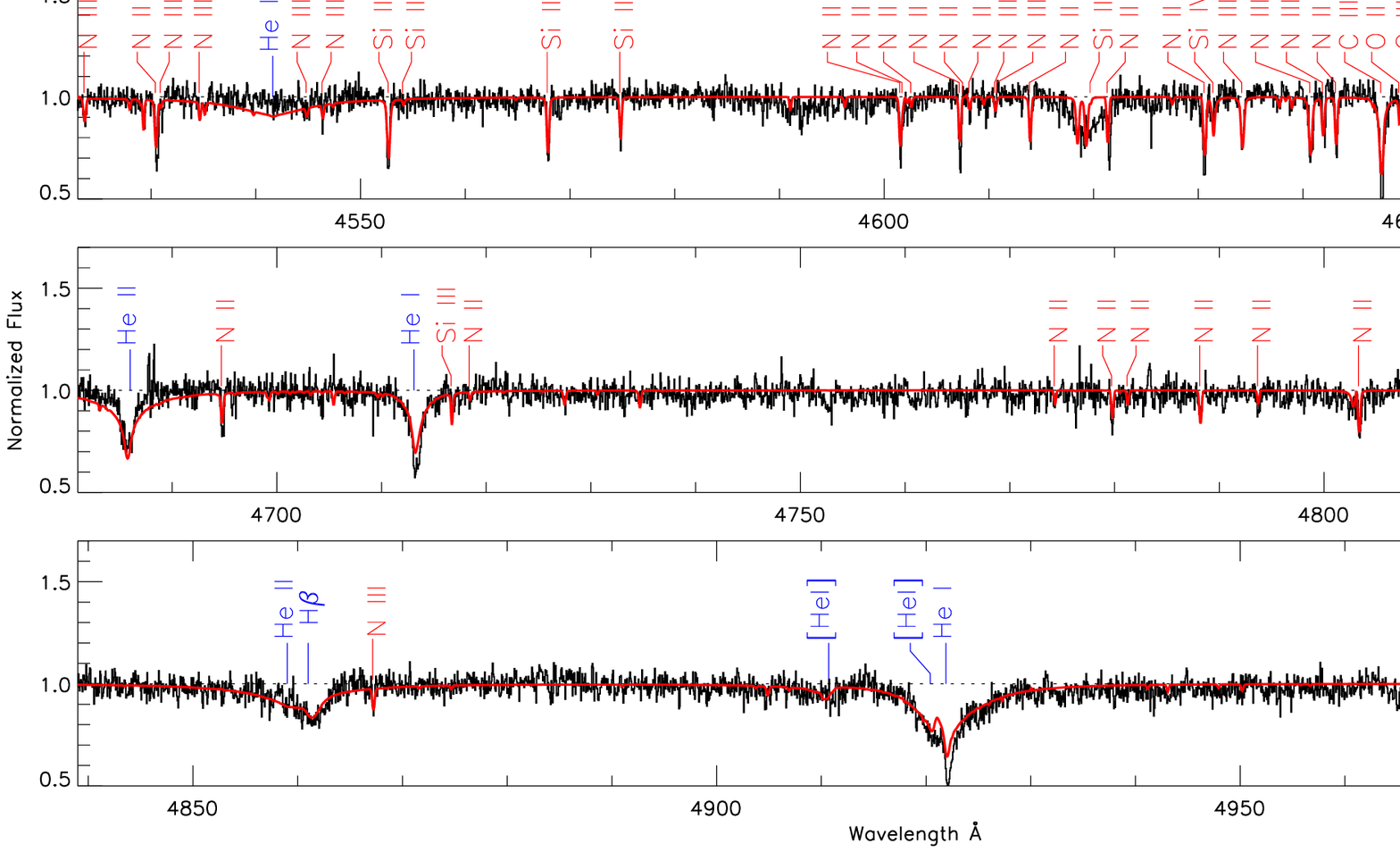}
\caption{The AAT spectrum of BPS\,22940--0009 along with the best-fit model
  of $T_{\rm eff}=34\,000\,{\rm K}$, $\log g=4.5$, $n_{\rm He}=0.989$, $v
  \sin i=4\,{\rm km\,s^{-1}}$, and $v_{\rm t}=10\,{\rm km\,s^{-1}}$. Abundances are as in Table~\ref{t_abs}.  }
\label{f_BPS22940}
\end{figure*}

\begin{figure*}
\centering
\includegraphics[trim=0cm 3cm 0cm 0cm, height=9.0in,width=7.5in]{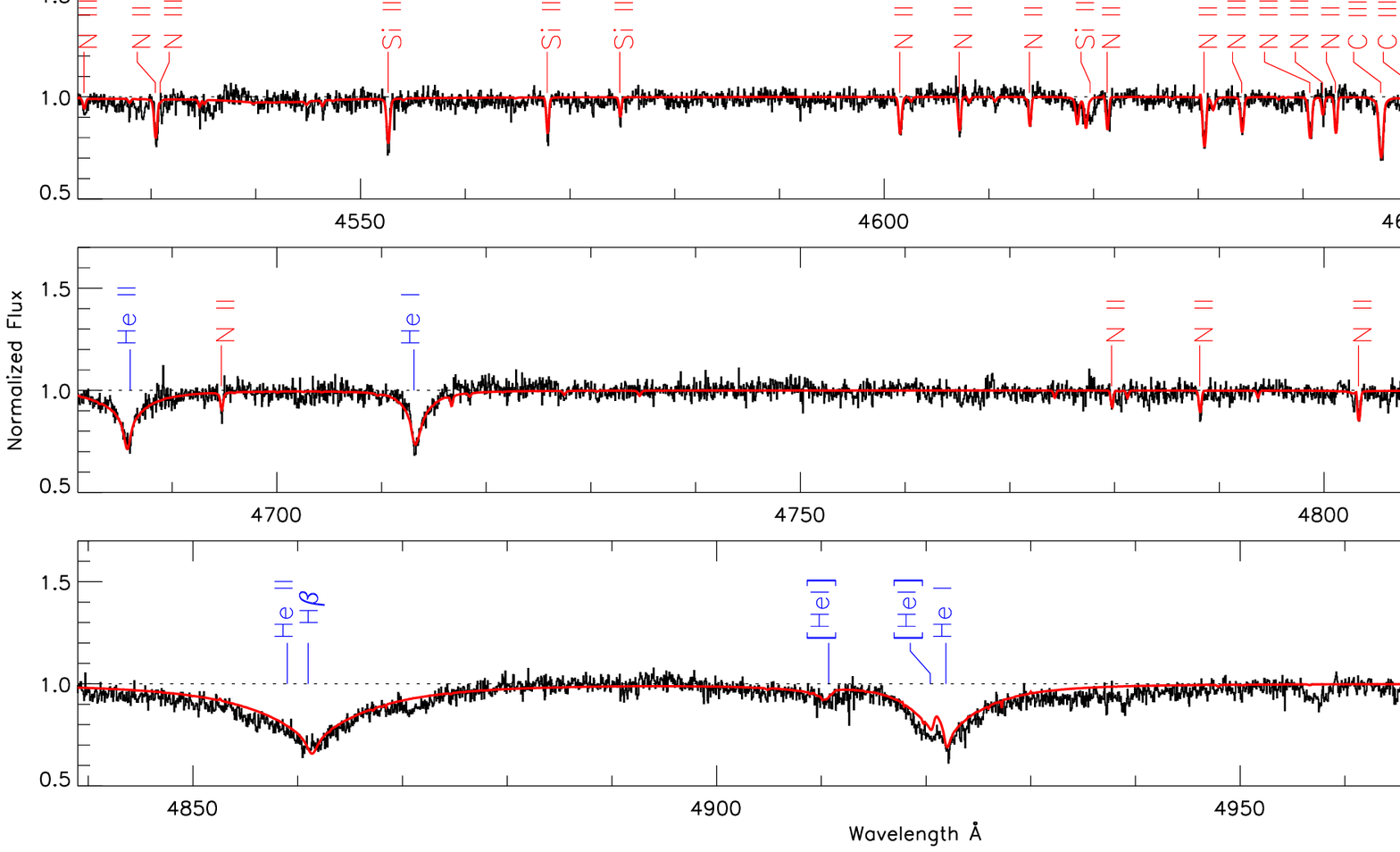}
\caption{The AAT spectrum of BPS\,22956--0094 along with the best-fit model
  of $T_{\rm eff}=34\,000\,{\rm K}$, $\log g=5.5$, $n_{\rm He}=0.699$, $v
  \sin i=2\,{\rm km\,s^{-1}}$, and $v_{\rm t}=5\,{\rm km\,s^{-1}}$. Abundances are as in Table~\ref{t_abs}. }
\label{f_BPS22956}
\end{figure*}

\begin{figure*}
\centering
\includegraphics[trim=0cm 3cm 0cm 0cm, height=9.0in,width=7.5in]{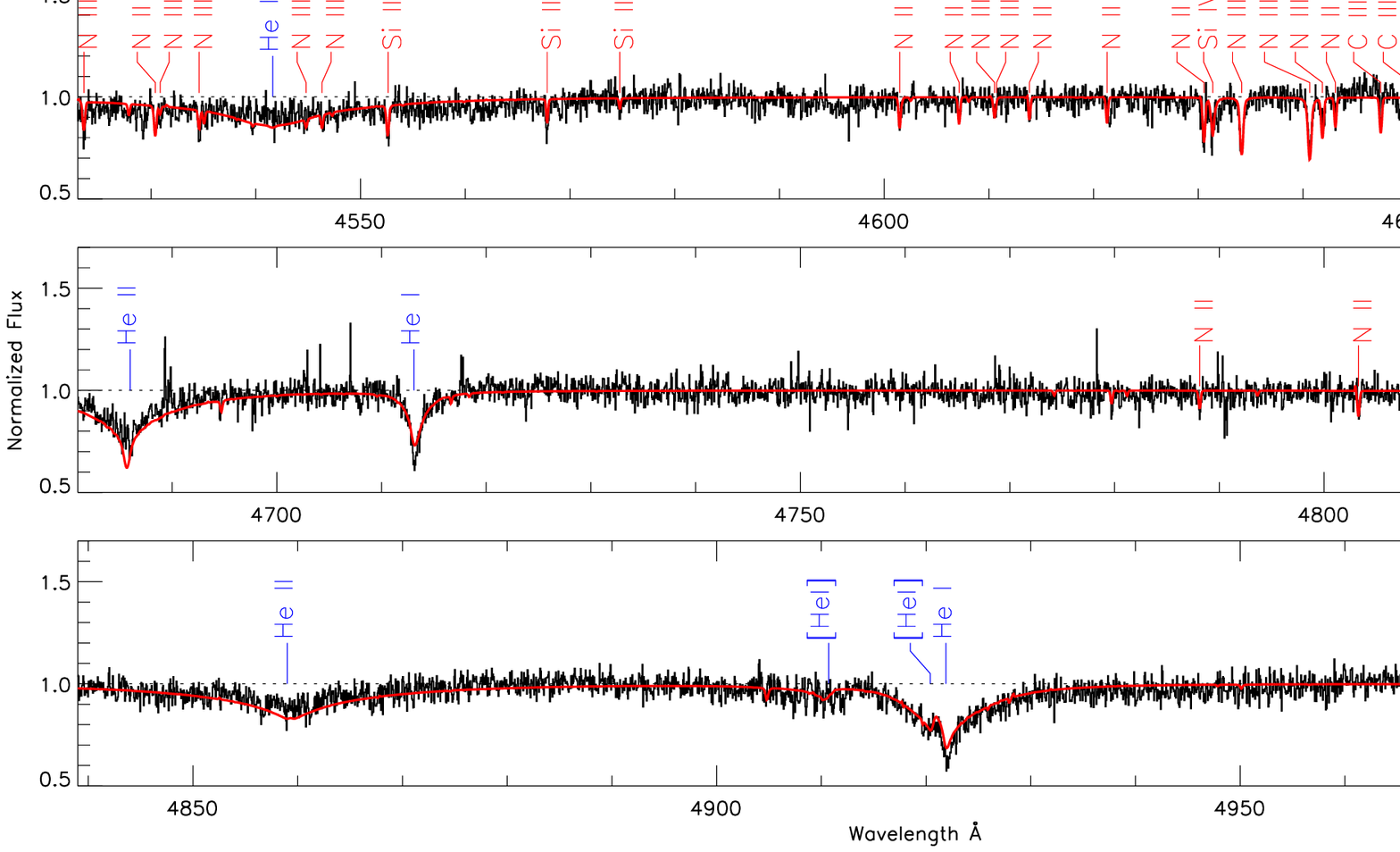}
\caption{The AAT spectrum of BPS\,29496--0010 along with the best-fit model
  of $T_{\rm eff}=40\,000\,{\rm K}$, $\log g=5.5$, $n_{\rm He}=0.999$, $v
  \sin i=2\,{\rm km\,s^{-1}}$, and $v_{\rm t}=10\,{\rm km\,s^{-1}}$. Abundances are as in Table~\ref{t_abs}. }
\label{f_BPS29496}
\end{figure*}

\end{document}